\documentclass[11pt, a4paper]{article}

\usepackage{amsthm}
\usepackage{amsmath}
\usepackage{amssymb}
\usepackage{amsfonts}
\usepackage{latexsym}
\usepackage{booktabs}
\usepackage{multirow}
\usepackage{graphicx}
\usepackage{natbib}
\usepackage{pdfpages}
\usepackage{algorithm}
\usepackage{algpseudocode}
\usepackage{bm}
\usepackage[left = 2.5cm, right = 2.5cm, top = 3cm, bottom = 3cm]{geometry}
\usepackage[colorlinks,citecolor=black,urlcolor=black]{hyperref}

\newtheoremstyle{example}
{3pt} 
{3pt} 
{} 
{0\parindent} 
{\bf}
{:} 
{.5em} 
{} 
\newtheoremstyle{theorem}
{3pt} 
{3pt} 
{\em} 
{0\parindent} 
{\bf}
{:} 
{.5em} 
{} 
\theoremstyle{example} \newtheorem{example}{Example}[section]
\theoremstyle{theorem}

\def\btheta{\theta}
\def\blambda{\lambda}
\def\bbb{b}
\def\bi{i}
\def\bA{A}
\def\bbeta{\beta}
\def\bgamma{\gamma}
\def\bX{X}
\def\by{y}
\def\bW{W}

\def\bR{R}
\def\bL{L}
\def\bT{T}
\def\bx{x}
\def\b1{1}

\title{Location-adjusted Wald statistics for scalar parameters}

\author{Claudia Di Caterina \\ \texttt{dicaterina@stat.unipd.it}
	\smallskip \\
  Department of Statistical Sciences, University of Padova \\ Via
  Cesare Battisti 241, 35121 Padova, Italy \\ \bigskip \\
  Ioannis Kosmidis
  \hspace{.2cm}\\
  \texttt{Ioannis.Kosmidis@warwick.ac.uk} \smallskip \\
Department of Statistics \\
University of Warwick \\
Coventry, CV4 7AL, UK \smallskip \smallskip \\
The Alan Turing Institute \\
96 Euston Road, London NW1 2DB, UK}

\begin{document}

\maketitle

\begin{abstract}
	\noindent
 Inference about a scalar parameter of interest is a core statistical
 task that has attracted immense research in statistics. The Wald
 statistic is a prime candidate for the task, on the grounds of the
 asymptotic validity of the standard normal approximation to its
 finite-sample distribution, simplicity and low computational cost.
 It is well known, though, that this normal approximation can be
 inadequate, especially when the sample size is small or moderate
 relative to the number of parameters. A novel, algebraic
 adjustment to the Wald statistic is proposed, delivering significant
 improvements in inferential performance with only small
 implementation and computational overhead, predominantly due to
 additional matrix multiplications. The Wald statistic is viewed as
 an estimate of a transformation of the model parameters and is
 appropriately adjusted, using either maximum likelihood or
 reduced-bias estimators, bringing its expectation asymptotically
 closer to zero. The location adjustment depends on the expected
 information, an approximation to the bias of the estimator, and the
 derivatives of the transformation, which are all either readily
 available or easily obtainable in standard software for a wealth of
 models. An algorithm for the implementation of the location-adjusted
 Wald statistics in general models is provided, as well as a
 bootstrap scheme for the further scale correction of the
 location-adjusted statistic. Ample analytical and numerical evidence is 
 presented for the adoption of the location-adjusted statistic in 
 prominent modelling settings, including inference about log-odds and 
 binomial proportions, logistic regression in the presence of
 nuisance parameters, beta regression, and gamma regression. The
 location-adjusted Wald statistics are used for the construction of
 significance maps for the analysis of multiple sclerosis lesions
 from MRI data.  \smallskip


\end{abstract}


\section{Introduction}
\label{sec:intro}

Testing hypotheses and constructing confidence intervals for scalar
parameters are key statistical tasks that are usually carried out
relying on large-sample results about likelihood-based
quantities. Under a model that is specified partially or fully by the
null hypothesis, the signed root of the logarithm of the likelihood
ratio (LR) statistic, the score statistic and the Wald statistic are
equivalent to first-order \citep[\S~3.4.1]{ps97}, yet the use of the
latter for inference is more widespread. The Wald statistic involves a
direct comparison between the estimated and the hypothetical value of
the parameter, accounting also for estimation uncertainty. As a
result, and in contrast to its main competitors, its computation does
not require fitting the model under the null hypothesis, which can be
time-consuming for complex models or when there is a need to perform
many tests.

The Wald test can, though, demonstrate anomalies in its power, mainly
because of the use of parameter estimates in the variance part of the
statistic \citep{mantel}. \citet{haudon} and \citet{fearsetal} study
such anomalies in logistic regression models and in one-way random
effects analyses of variance, respectively, and \citet{vaeth} gives
mathematical conditions under which Wald procedures can suffer from
scarce power in the more general context of exponential family
models. In addition, the performance of Wald-type inference depends
directly and, sometimes, critically on the properties of the estimator
used in the statistic. A common strategy in enhancing Wald inference
is to replace the maximum likelihood (ML) estimator with another that
has improved frequentist properties and a limiting normal
distribution. One prominent example is the use of a moment-based
estimator for the dispersion parameter in generalized linear models
with unknown dispersion, as is recommended in
\citet[\S~8.3]{mcculnel89} and implemented in the \texttt{summary.glm}
function of the \texttt{stats} R package \citep{R}. Another recent
example is in \citet{kosfir10}, who illustrate that the finite-sample
bias of the ML estimator of the precision parameter in beta regression
models results in excessively narrow Wald-type confidence intervals (CIs)
and anti-conservative Wald tests, and propose the use of a
reduced-bias (RB) estimator to alleviate those issues. Below is a
working illustration of that proposal in a beta regression model
involving precision covariates.

\begin{example}
  \label{reading_skills}
  \citet{smiver06} use beta regression to investigate the relative
  contribution of nonverbal IQ to the distribution of $n = 44$
  children's scores on a reading accuracy test, controlling for the
  presence of diagnosed dyslexia. The score of the $i$th child is
  assumed to be from a beta random variable with mean $\mu_i$ and
  variance $\mu_i(1- \mu_i)/(1 + \phi_i)$, with $\phi_i > 0$. The
  score mean $\mu_i$ and precision $\phi_i$ are linked with covariates
  through the relationships
  \begin{equation}
    \label{eq:m}
    \log\frac{\mu_i}{1 - \mu_i} = \beta_1 + \sum_{j=2}^{4}\beta_{j}x_{ij}
    \quad \text{and} \quad
    \log\phi_i = \gamma_1 + \sum_{j=2}^{3}\gamma_{j}x_{ij} \quad
    (i = 1, \dots, n)\,,
  \end{equation}
  where
  $\bbeta = (\beta_1, \beta_2, \beta_3, \beta_4)^\top \in \Re^4,
  \bgamma = (\gamma_1, \gamma_2, \gamma_3)^\top \in \Re^3$ are the
  vectors of regression coefficients, $x_{i2}$ takes value $-1$ if the
  $i$th child is dyslexic and $1$ if not, $x_{i3}$ is the nonverbal IQ
  score, and $x_{i4} = x_{i2}x_{i3}$ is the interaction between
  dyslexia status and nonverbal IQ score.

  \begin{table}[t]
    \caption{Maximum likelihood (ML) and reduced-bias (RB)
      estimates, corresponding standard errors (in parenthesis) and
      $95\%$ Wald-type confidence intervals for the parameters of
      model~(\ref{eq:m}).}
    \begin{center}
      \begin{small}
      \begin{tabular}{lrrrrrrrr}
        \toprule
        & \multicolumn{4}{c}{Estimates} & \multicolumn{4}{c}{95\% Confidence intervals} \\
        & \multicolumn{2}{c}{ML} & \multicolumn{2}{c}{RB} & \multicolumn{2}{c}{ML} & \multicolumn{2}{c}{RB} \\
        \midrule
        $\beta_1$ & 1.123 & (0.143) & 1.114 &  (0.148) & 0.843 & 1.403 & 0.824 & 1.405 \\
        $\beta_2$ & -0.742 & (0.143) & -0.734 & (0.148) & -1.021 & -0.462 & -1.024 & -0.444\\
        $\beta_3$ & 0.486 & (0.133) & 0.441 & (0.141) & 0.225 & 0.747 & 0.165 & 0.717\\
        $\beta_4$ & -0.581 & (0.133) & -0.532 & (0.140) & -0.841 & -0.321 & -0.807 & -0.257\\
        $\gamma_1$ & 3.304 & (0.223) & 3.092 & (0.225) & 2.868 & 3.741 & 2.652 & 3.533\\
        $\gamma_2$ & 1.747 & (0.262) & 1.654 & (0.264) & 1.232 & 2.261 & 1.138 & 2.171\\
        $\gamma_3$ & 1.229 & (0.267) & 1.048 & (0.271) & 0.705 & 1.753 & 0.518 & 1.578\\
        \bottomrule

      \end{tabular}
    \end{small}
    \label{tab:fit1}
    \end{center}
  \end{table}

  The parameters $\bbeta$ and $\bgamma$ in (\ref{eq:m}) are estimated
  using the ML estimator and its RB version, as these are
  implemented in the R package \texttt{betareg}
  \citep{grun}. Table~\ref{tab:fit1} shows the resulting estimates,
  the corresponding estimated standard errors based on the expected
  information matrix, and the nominally 95\% individual Wald-type
  CIs. As is also noted in \citet{kosfir10}, bias
  reduction inflates estimated standard errors, and results in
  CIs that are wider than those based on ML, better
  reflecting the uncertainty about the values of the parameter. To
  illustrate this, the coverage probabilities of individual Wald-type
  intervals are estimated at levels $90\%$, $95\%$ and $99\%$, using
  $50\, 000$ samples simulated under the ML fit. The results in
  Table~\ref{covCI} suggest that the use of the RB estimates
  in Wald-type CIs brings the empirical coverage
  probabilities closer to the nominal value.

  \begin{table}[t]
    \caption{Empirical coverage probabilities of individual Wald-type
      confidence intervals for
      $\beta_2, \beta_3, \beta_4, \gamma_2, \gamma_3$ in~(\ref{eq:m}),
      based on the maximum likelihood estimator (ML) and its
      reduced-bias version (RB) at nominal levels 90\%, 95\%, and
      99\%. Reported rates based on $50\, 000$ samples simulated under
      the ML fit in Table~\ref{tab:fit1}.}
    \begin{center}
      \begin{small}
        \begin{tabular}{lcccccc}
        \toprule
        & \multicolumn{3}{c}{ML} & \multicolumn{3}{c}{RB}
        \\ \cmidrule{2-7}
        & $90\%$ & $95\%$ & $99\%$ & $90\%$ & $95\%$ & $99\%$ \\ \midrule
        $\beta_2$ & 86.9 & 92.4 & 97.7 & 88.1 & 93.4 & 98.2 \\
        $\beta_3$ & 84.8 & 91.0 & 97.1 & 87.2 & 92.9 & 98.0 \\
        $\beta_4$ & 85.0 & 91.2 & 97.2 & 87.3 & 92.9 & 98.0 \\
        $\gamma_2$ & 82.4 & 89.1 & 96.1 & 83.8 & 90.2 & 96.7 \\
        $\gamma_3$ & 79.1 & 86.0 & 94.4 & 82.7 & 89.2 & 96.1  \\ \bottomrule
      \end{tabular}
    \end{small}
  \end{center}
    \label{covCI}
  \end{table}
\end{example}

In the current paper, we improve Wald inference about scalar
parameters through a novel, more explicit approach than the one
presented in Example~\ref{reading_skills}, which exploits the direct
dependence of the Wald statistic on the estimator. Location-adjusted (LA)
Wald statistics are defined whose expectations are asymptotically
closer to that of the limiting normal distribution under the null
hypothesis. The developments in this paper also shed light on why the
use of RB estimators in the Wald statistics can improve
inferential performance in cases, like it does in
Example~\ref{reading_skills}.

Corresponding methods have been first proposed by \citet{bartlett37}
for enhancing first-order inference based on the LR
statistic, and have later been applied to other test statistics
\citep[see, for example,][]{corfer91}. The location adjustment that is
introduced here depends on quantities that are either readily
available or easily obtainable, either analytically or numerically,
for many well-used model classes. In particular, the proposed
adjustment involves the expected information, an approximation to the
bias of the estimator \citep[see, for example][for the first-term in
the bias expansion of the ML estimator]{coxsne68}, and the derivatives
of an appropriate transformation of the model parameters, which can be
computed either analytically or by numerical or automatic
differentiation methods.

We demonstrate how the correction in the location of the Wald
statistic strikes a balance in being sufficient to deliver significant
improvements to finite-sample inferential performance over other
status-quo methods in prominent modelling scenarios, only with a small
sacrifice to the computational simplicity of classical Wald inference,
mainly due to extra matrix multiplications. A bootstrap procedure that
exploits the computational simplicity of LA Wald statistics
is also presented to deliver location- and scale-adjusted Wald statistics.

Our recommendation of adopting LA Wald statistics in statistical
practice is supported by ample analytical and numerical evidence,
along with case-studies under well-used modelling settings. These
settings include inference about log-odds and binomial proportions,
inference from logistic regression models in the presence of nuisance
parameters, beta regression, gamma regression and random-effects
meta-analysis (see, supplementary material) models.
The LA Wald statistics are successfully used within the
mass univariate probit regression framework in
\citet[Subsection~4.1]{ge2014} for the construction of brain
significance maps to visualize the strength of association of patient
characteristics to the occurrence of multiple sclerosis lesions from
MRI data. This is a scenario where standard statistics, like the Wald
and likelihood-ratio ones, have sub-optimal properties or completely
fail to apply due to the occurrence of infinite estimates in many of
the thousands voxel-wise regressions.

The current paper is structured as follows. The location adjustment to
the standard Wald statistic and the Wald statistic based on
RB estimators are introduced in Section~\ref{sec:regset} and
Section~\ref{sec:RBest}. Section~\ref{sec:comp} derives the
computational complexity of the LA Wald statistics and
presents an algorithm for their computation for general
models. Section~\ref{sec:exp} illustrates the effect that the location
adjustment has on the normal approximation to the distribution of the
Wald statistic. Procedures for the computation of CIs
are detailed in Section~\ref{sec:confint}, and their performance is
assessed under a beta regression setting in
Section~\ref{sec:beta}. Section~\ref{sec:bernoulli} gives evidence
on the accuracy of the LA Wald statistics when making
inference on log-odds and binomial proportions. Section~\ref{sec:glm}
obtains the closed form of the quantities required to compute
LA Wald statistics for inference from generalized
linear models, and shows simulation results for gamma and logistic
regressions. In Section~\ref{sec:MRI}, the LA
statistics are used for the construction of significance maps for the
analysis of multiple sclerosis lesions from MRI data. Finally,
Section~\ref{sec:scale} presents a bootstrap procedure for the
scale-adjustment of LA Wald statistics, before closing
with discussion and further work in Section~\ref{sec:conc}.

\section{Location-adjusted Wald statistic}
\label{sec:regset}

\subsection{Bias of the Wald statistic}
Consider a sample $\by =(y_1,\dots,y_n)^\top$ of observations assumed
to be realizations of independent random variables $Y_1,\dots,Y_n$,
with $Y_i$ having conditional density or probability mass function
$f(y_i | \bx_i; \btheta)$, where $\btheta\in \Theta \subseteq \Re^p$,
$p \le n$, is the parameter vector and
$\bx_i=(x_{i1}, \dots, x_{ik})^\top$ is a $k$-vector of explanatory
variables for the $i$th observation $(i=1,\dots,n)$. We partition
$\btheta$ as $\btheta = (\psi, \blambda^\top)^\top$, where
$\psi \in \Psi \subset \Re$ is a scalar parameter of interest and
$\blambda \in \Lambda \subset \Re^{p-1}$ is a $(p - 1)$-vector of
nuisance parameters.

Assuming that the log-likelihood function
$l(\btheta)=\sum_{i=1}^{n}\log f(y_i | \bx_i; \btheta)$ satisfies the
usual regularity conditions \citep[see, for example,][\S~3.4]{ps97}, a
typical way to construct inference about $\psi$ is using a Wald
statistic. For example, the Wald test for $H_0\!:\psi=\psi_0$ with
$\psi_0 \in \Re$ computes $p$-values using the standard normal
distribution and the observed value of the signed Wald statistic
\begin{equation}
  \label{reg:t}
  t = (\hat\psi-\psi_0)/\kappa(\hat{\btheta})\,.
\end{equation}
In the above expression,
$\hat{\btheta}=(\hat{\psi}, \hat{\blambda}^\top)^\top=\arg \max_{\btheta
  \in \Theta} l(\btheta)$ is the ML estimate of
$\btheta$ and $\kappa(\btheta)$ is the square root of the
$(\psi, \psi)$-element of the variance-covariance matrix
$\{\bi(\btheta)\}^{-1}$ of the exact or asymptotic distribution of the
estimator $\hat\btheta$, usually taken as the inverse of the expected
information $E\{ \nabla l(\btheta) \nabla l(\btheta)^\top \}$, where $\nabla$
denotes the gradient with respect to $\btheta$. Without
loss of generality, we assume that the element at the first row and
first column of $\{\bi(\btheta)\}^{-1}$ is the asymptotic variance of
$\hat\psi$.

The standard normal distribution is not always a good approximation to
the exact distribution of (\ref{reg:t}) under $H_0$. This is commonly
the case when the model is highly non-linear in the parameters or $n$
is small or moderate relative to $p$ \citep[see,
for example,][\S~6.2.4]{mcculnel89}.

We show how this approximation can be easily improved by bringing the
first null moment of the asymptotic distribution for the adjusted Wald
statistic closer to zero. The word ``null'' here is used to highlight
the fact that the expectation is taken with respect to the model,
assuming that the null hypothesis holds.

Consider the transformation
\begin{equation}
  \label{testf}
  T(\btheta; \psi_0) = (\psi-\psi_0)/\kappa(\btheta)
\end{equation}
of the parameter $\btheta$. We call $T(\btheta; \psi_0)$ the Wald
transform. The Wald transform is of order $O\big(n^{1/2}\big)$ because
$\kappa(\btheta) = O\big(n^{-1/2}\big)$. Then, $t$ in (\ref{reg:t}) is
the ML estimator of (\ref{testf}). As is the case for
$\hat{\btheta}$, $t$ is also subject to finite-sample bias, which can
be reduced by subtracting the first term in the asymptotic expansion
of its bias, as is shown, for example, in \citet[Remark~11,
p.~1214]{efron75}.

Assume that (\ref{testf}) is at least three times differentiable with
respect to $\btheta$ and that $\hat\btheta$ is consistent. Then, using
the Einstein summation convention,
$T(\hat{\theta}; \psi_0) - T(\theta; \psi_0)$ can be expanded as
\begin{align}
  \label{biast1}
  (\hat{\theta}^u -\theta^u)T_u(\theta; \psi_0)
  & + \frac{1}{2}(\hat{\theta}^u - \theta^u)(\hat{\theta}^v -
    \theta^v) T_{uv}(\theta; \psi_0) \\
  & + \frac{1}{6}(\hat{\theta}^u - \theta^u)(\hat{\theta}^v -
    \theta^v)(\hat{\theta}^w - \theta^w)T_{uvw}(\theta; \psi_0)
    + O_p\big(n^{-3/2}\big)\,,\nonumber
\end{align}
where $T_u(\theta; \psi_0), T_{uv}(\theta; \psi_0)$ and
$T_{uvw}(\theta; \psi_0)$ are the gradient, hessian and third
derivative, respectively, of (\ref{testf}) $(u, v, w = 1, \dots, p)$,
all with order $O\big(n^{1/2}\big)$. Taking expectations in
(\ref{biast1}) gives that
$E\{T(\hat{\theta}; \psi_0) - T(\theta; \psi_0)\} = B(\theta; \psi_0)+
O\big(n^{-3/2}\big)$ with
\begin{equation}
  \label{biastj_index}
  B(\theta; \psi_0) = b^u(\theta)T_u(\theta; \psi_0) +
  \frac{1}{2}i^{u,v}(\theta)T_{uv}(\theta;
  \psi_0) \,,
\end{equation}
where the first-order bias $b^u(\theta)$ is such that
$E_{\theta}(\hat{\theta}^u - \theta^u)=b^u(\theta) +
o\big(n^{-1}\big)$ and $i^{u,v}(\theta)$ can be understood as the
$(u,v)$th element of $\{\bi(\btheta)\}^{-1}$ $(u,v=1,\dots,p)$.  The
above expression can also be derived using \citet[\S~4.3,
Remark~3]{kosfir10}, which gives the first-term in the bias expansion
of the ML estimator for a transformation of a scalar
parameter in terms of that of the ML estimator for the
parameter. Expression~(\ref{biastj_index}) can be written in matrix
notation as
\begin{equation}
  \label{biastj}
  B(\btheta, \psi_0) = \{\bbb(\btheta)\}^\top \nabla T(\btheta; \psi_0) +
  \frac{1}{2} \mathrm{tr} \left[ \{\bi(\btheta)\}^{-1} \nabla \nabla^\top
    T(\btheta; \psi_0) \right]
  \, ,
\end{equation}
where $\mathrm{tr}(\bA)$ denotes the trace of matrix $\bA$, and $\nabla$
and $\nabla\nabla^\top$ denote the gradient and the matrix of second
derivatives with respect to $\btheta$, respectively.

\subsection{Location-adjusted Wald statistic}

The LA Wald statistic for $\psi$ is then
\begin{equation}
  \label{reg:t*}
  t^{*}= t - \widehat B\,,
\end{equation}
where $\widehat B$ is a suitable estimator of $B(\btheta; \psi_0)$
in~(\ref{biastj}). Natural candidates for $\widehat B$ are
$B(\hat{\btheta}_0; \psi_0)$ and $B(\hat{\btheta}; \psi_0)$, where
$\hat{\btheta}_0 = (\psi_0, \hat\blambda_0^\top)^\top$ and
$\hat\blambda_0 = \arg \max_{\blambda \in \Lambda} l(\psi_0, \blambda)$
is the constrained ML estimate of $\blambda$. In either
case, a calculation along the lines of \citet[\S~9.42]{ps97} shows
that the null expectation of $t^*$ is $O(n^{-3/2})$, that is
asymptotically closer to zero than the null expectation of $t$, which
is $O(n^{-1/2})$.

We focus on LA statistics with
$\hat{B} = B(\hat{\btheta}; \psi_0)$ because their computation does not
require any additional constrained or otherwise optimization, than the
already available fit for the full model.

\subsection{Derivatives of the Wald transform}

Expression~(\ref{biastj}) is convenient because it depends only on the
derivatives of $T(\btheta; \psi_0)$, the elements of
$\{\bi(\btheta)\}^{-1}$, and the first term in the expansion of the
bias of the ML estimator, which has been derived in
\citet[expression~(20)]{coxsne68} for general parametric models, and is
given in matrix form in \citet[Section~2]{kosfir10}.

Both $\bi(\btheta)$ and the first-order bias are
readily available for a wide range of well-used model classes,
especially those for which asymptotic bias reduction methods have been
implemented \citep[see, among others,][]{cook86, cordmc91, cordeiro97,
  botter98, simas, cordeiro08, grun}. In addition, the derivatives of
the Wald transform can be written in terms of derivatives of
$\kappa(\btheta)$. Specifically, the gradient
$\nabla T(\btheta; \psi_0)$ and the hessian
$\nabla\nabla^\top T(\btheta; \psi_0)$ are
\begin{equation}
  \label{der1:t}
  \left\{ e_{p} - T(\btheta;
    \psi_0) \nabla \kappa(\btheta)\right\}/\kappa(\btheta)
\end{equation}
and
\begin{equation}
  \label{der2:t}
  -\left[
    \nabla \kappa(\btheta) \left\{\nabla
      T(\btheta; \psi_0) \right\}^\top +
    \nabla T(\btheta; \psi_0) \left\{\nabla
      \kappa(\btheta)\right\}^\top +
    T(\btheta; \psi_0) \nabla \nabla^\top \kappa(\btheta)
  \right]/\kappa(\btheta) \, ,
\end{equation}
respectively. In the above expressions, $ e_{p}$ is a $p$-vector with
first element one and zeros everywhere else.


Expressions~(\ref{der1:t}) and (\ref{der2:t}) can, in turn, be written
in terms of the information matrix $\bi(\btheta)$ and its derivatives. A
straightforward application of matrix differentiation rules
\citep[see, for example][]{magnus99} gives that the generic $u$th
element of the $p$-dimensional gradient vector $\nabla \kappa(\btheta)$
has the form
\begin{equation}
  \label{kappa_gradient}
  \frac{\partial \kappa(\btheta)}{\partial \theta_u} = -
  \frac{1}{2\kappa(\btheta)}\bigg[ \{\bi(\btheta) \}^{-1} \frac{\partial
    \bi(\btheta)}{\partial \theta_u}  \{\bi(\btheta) \}^{-1}\bigg]_{\psi\psi}
  \qquad (u = 1, \ldots, p) \, ,
\end{equation}
where $[\,\cdot\,]_{\psi\psi}$ denotes the $(\psi, \psi)$ element of the
square matrix in brackets. Further differentiation gives that
the $(u,v)$th element in the $p \times p$ hessian
$\nabla \nabla^\top \kappa(\btheta)$ is
\begin{align}
  \label{kappa_hessian}
  \frac{\partial^2 \kappa(\btheta)}{\partial \theta_u \partial \theta_v} = -&\frac{1}{4\{\kappa(\btheta)\}^3}\bigg[ \{\bi(\btheta) \}^{-1} \frac{\partial \bi(\btheta)}{\partial \theta_u} \{\bi(\btheta) \}^{-1}\bigg]_{\psi\psi}
                                                                                \bigg[
                                                                             \{\bi(\btheta)
                                                                             \}^{-1}
                                                                             \frac{\partial
                                                                             \bi(\btheta)}{\partial
                                                                             \theta_v}
                                                                             \{\bi(\btheta)
                                                                             \}^{-1}\bigg]_{\psi\psi}
  \\ \notag
  +& \frac{1}{2\kappa(\btheta)}\bigg[ \{\bi(\btheta) \}^{-1} \frac{\partial \bi(\btheta)}{\partial \theta_v} \{\bi(\btheta) \}^{-1} \frac{\partial \bi(\btheta)}{\partial \theta_u} \{\bi(\btheta) \}^{-1}\bigg]_{\psi\psi}
     - \frac{1}{2\kappa(\btheta)}\bigg[ \{\bi(\btheta) \}^{-1}
     \frac{\partial^2 \bi(\btheta)}{\partial \theta_u \partial \theta_v}  \{\bi(\btheta) \}^{-1}\bigg]_{\psi\psi} \\ \notag
  +& \frac{1}{2\kappa(\btheta)}\bigg[ \{\bi(\btheta) \}^{-1} \frac{\partial
     \bi(\btheta)}{\partial \theta_u} \{\bi(\btheta) \}^{-1} \frac{\partial
     \bi(\btheta)}{\partial \theta_v} \{\bi(\btheta) \}^{-1}\bigg]_{\psi\psi}
     \qquad (u, v = 1, \ldots, p)  \, . \notag
\end{align}

The information matrix $\bi(\btheta)$ and its derivatives are
model-dependent, but can be found directly for popular model
classes, as done in Subsection~\ref{sec:computation_glms} for
generalized linear models.

\section{Wald statistics and reduced-bias estimators}
\label{sec:RBest}
As is illustrated in Example~\ref{reading_skills} and the works cited
therein, the use of RB estimators can deliver marked
improvements in the inferential performance of Wald-type
procedures. These performance gains can be partly explained by the
fact that employing asymptotically efficient estimators with bias of
order $o(n^{-1})$ in the Wald statistic results in the bias of the
latter being as in~(\ref{biastj}) but without the term
$\{\bbb(\btheta)\}^\top \nabla T(\btheta; \psi_0)$. To see this,
notice that $T(\tilde{\btheta}; \psi_0) - T(\btheta; \psi_0)$ admits
the same expansion as in~(\ref{biast1}), with $\hat{\btheta}$ replaced
by the RB estimator $\tilde\btheta$. The expectation of the
term $(\tilde\theta^u - \theta^u) T_u (\theta; \psi_0)$ is
$o(n^{-1/2})$, though, giving
$E\{T(\tilde{\btheta}; \psi_0) - T(\btheta; \psi_0)\} =
\tilde{B}(\btheta; \psi_0) + o(n^{-1/2})$, where, in matrix notation
\begin{equation}
  \label{biast*br}
  \tilde{B}(\btheta; \psi_0) = \frac{1}{2} \mathrm{tr} \left[ \{\bi(\btheta)\}^{-1} \nabla \nabla^\top
    T(\btheta; \psi_0) \right]\,.
\end{equation}
The above quantity is of the same order, $O(n^{-1/2})$, as
(\ref{biastj}). Given that the signs and the magnitude of terms in the
right hand side of~(\ref{biastj}) are rarely known for general models,
the elimination of $\{\bbb(\btheta)\}^\top \nabla T(\btheta; \psi_0)$
through the use of RB estimators provides no guarantee of
improvement for general models, despite the positive results in the
setting of Example~\ref{reading_skills}. In fact, employment of RB
estimators in the Wald statistics may even have a detrimental effect
on inference.

Expression~(\ref{biast*br}) proves useful, though, for adjusting the
location of Wald statistics based on RB estimators, in settings where
the latter have been found to be particularly desirable, especially
with reference to Wald-type inference. Such a scenario, which is also
popular in recent applied work, involves models for categorical data
where RB estimators are always finite, even in cases where the ML
estimates have infinite components \citep[see, for example,][for
logistic regression and regression models with nominal and ordinal
responses, respectively]{heinze02, kosfir11, kosmidis14b}. Despite
that the Wald statistic and its LA version cannot be directly computed
when $\hat\psi$ is infinite, the Wald statistic based on the RB
estimator $\tilde{t} = T(\tilde\btheta; \psi_0)$ and its LA version
\[
  \tilde{t}^* = \tilde{t} - \tilde{B}(\tilde{\btheta};  \psi_0) \, ,
\]
are both well-defined. By the same arguments to those in
Section~\ref{sec:regset} for the development of $t^*$, the null
expectation of $\tilde{t}^*$ is $O(n^{-3/2})$ and, hence,
asymptotically closer to zero than that of $\tilde{t}$.

\section{Implementation and computational complexity}
\label{sec:comp}

Assuming that the estimates for $\btheta$ are available and that the
information $\bi(\btheta)$ has been computed at those estimates, the
inversion of $\bi(\btheta)$ involves, typically, $O(p^3)$
operations. Hence, the complexity for the computation of the Wald
statistic~(\ref{reg:t}) is $O(p^3)$.

After having computed the derivatives of $\bi(\btheta)$ and
$\bbb(\btheta)$ at the estimates, and without exploiting any model
structures or sparsity in $\bi(\btheta)$, the computation of
$B(\hat{\btheta}; \psi_0)$ requires $O(p^{4})$ operations. This number
of operations is due to the matrix multiplications for computing the
gradient~(\ref{der1:t}), the hessian~(\ref{der2:t}) and, finally,
(\ref{biastj}). As a result, the LA Wald
statistic~(\ref{reg:t*}) based on $B(\hat{\btheta}; \psi_0)$ has,
generally, computational complexity $O(p^4)$. Note here that the extra
operations are just straightforward matrix multiplications and, hence,
the computing time can be significantly reduced by appropriate
vectorization, pre-computing some of the quantities, and parallelizing
others across the parameters, like, for instance, the products
$\{\bi(\btheta)\}^{-1} \partial \bi(\btheta) / \partial \theta_u$
across $u \in \{ 1, \ldots, p \}$.

\begin{algorithm}[t]
  \caption{Location-adjusted Wald statistics}
  \label{waldi}
\begin{algorithmic}[1]
\Procedure {LAWald}{$\btheta^*$, $\btheta_0$, $\bi$, $\bbb$, $c$, $e$}
\State $SE(\btheta, i) \leftarrow inverse(\bi(\btheta))[k, k]$
\Comment{$\kappa(\btheta)$}
\State $p \leftarrow length(\btheta^*)$
\Comment{length of the vector of estimates $\btheta^*$}
\If {$e = 1$}
\Comment{if $\btheta^*$ is ML estimate}
\State $B \leftarrow \bbb(\btheta^*)$
\EndIf
\Comment{$\bbb(\hat\btheta^*)$}
\State $I \leftarrow i(\btheta^*)$
\State $II \leftarrow inverse(I)$
\State $S \leftarrow vector(p)$
\Comment{$S$ as a $p$-vector}
\State $T \leftarrow vector(p)$
\Comment{$T$ as a $p$-vector}
\For {$j \in \{1, 2, \ldots, p\}$}
\State $S[j] \leftarrow Sqrt(II[j, j])$
\Comment{estimated standard error}
\State $T[j] \leftarrow (\btheta^*[j] - \btheta_0[j])/S[j]$
\Comment{Wald statistic}
\If {$c = 1$}
\Comment{if location-adjustment is requested}
\State $U \leftarrow numericgradient(SE, \btheta = \btheta^*, k = j)$
\Comment{$\nabla \kappa(\btheta)$ at $\btheta := \btheta^*$}
\State $V \leftarrow numerichessian(SE, \btheta = \btheta^*, k = j)$
\Comment{$\nabla\nabla^\top \kappa(\btheta)$ at $\btheta := \btheta^*$}
\State $A \leftarrow - T[j] * U$
\State $A[j] \leftarrow 1 + A[j]$
\If {$e = 1$}
\Comment{if $\btheta^*$ is ML estimate}
\State $W \leftarrow dotproduct(A, B)$
\Comment{dot product of $A$ and $B$}
\EndIf
\If {$e = 2$}
\Comment{if $\btheta^*$ is RB estimate}
\State $W \leftarrow 0$
\EndIf
\State $X \leftarrow wdotproduct(A, U, II)$
\Comment{dot product of $A$ and $U$ with weight-matrix $II$}
\State $Y \leftarrow sum(V * II)$
\Comment{sum of elements from element-wise product of $V$ and $II$}
\State $Y \leftarrow T[j] * Y$
\State $T[j] \leftarrow T[j] - (W + X + Y/2) / S[j]$
\Comment{location-adjusted Wald statistic}
\EndIf
\EndFor
\State \textbf{return} $T$
\EndProcedure
\end{algorithmic}
\end{algorithm}

If $p$ is not prohibitively large, and at the expense of additional
computing cost due to matrix inversions, the derivatives of
$\kappa(\btheta)$ in~(\ref{kappa_gradient}) and (\ref{kappa_hessian})
can be also calculated using numerical differentiation techniques at
$\hat\btheta$, provided there is an appropriate computer
implementation of the standard errors as a function of the
parameters. In this way, the location adjustment can be obtained for
general models, requiring only the ML or RB estimates, along with
ready implementations of the expected information matrix and an
approximation of the bias function. Algorithm~\ref{waldi} details such
a procedure in pseudo-code and the \texttt{waldi} R package
(\url{https://github.com/ikosmidis/waldi}) implements the computations
for generalized linear models and beta regression models. Notice that
Steps 12--29 of Algorithm~\ref{waldi} for calculating $t^*$ or
$\tilde{t}^*$ can be performed in parallel across parameters, with
significant savings in execution time when multiple computing units
are available. The \texttt{waldi} R package provides the parallel
computation of the LA Wald statistic.

\section{Effect of location adjustment on distributional approximation}
\label{sec:exp}

As shown in Section~\ref{sec:regset}, the location adjustment of the
Wald statistic delivers statistics with null expectations closer to
zero than the standard Wald statistic. This correction often extends
to higher-order moments than the mean, but it is not, overall,
sufficient to deliver a drop in the order of the error of the normal
approximation to the whole null distribution of the statistic, as
Bartlett-type corrections for the LR statistics do.

To see this, suppose that $Y_i$ has an exponential distribution with
mean $\mu_i=e^{-\theta}>0$, $\theta \in \Re$. The log-likelihood about
$\theta$ is $l(\theta)=n\theta-n\bar{y}e^\theta$, where
$\bar{y}=\sum_{i=1}^n y_i/n$ is the sample mean. The ML estimate, the
expected information and the first-order bias are
$\hat{\theta}=-\log\bar{y}$, $i(\theta) = n$, and
$b(\theta) = (2n)^{-1}$, respectively. The derivatives of the Wald
transform are $d T(\theta; \theta_{0})/ d\theta = n^{1/2}$ and
$d^2 T(\theta; \theta_{0})/d \theta^2 = 0$, and so
$B(\theta; \theta_0) = n^{-1/2}/2$. Hence, the Wald
statistic~(\ref{reg:t}) for $H_0\!: \theta = \theta_0$ is
$t = -n^{1/2}(\log \bar{y}+\theta_{0})$ and the LA Wald
statistic in (\ref{reg:t*}) is
$t^{*}= - n^{1/2}(\log \bar{y}+\theta_{0}) - n^{-1/2}/2$.

The Edgeworth expansions \citep[\S~2.3]{hall92} of the null
distribution functions $F(z)$ and $F^*(z)$ of $t$ and $t^{*}$,
respectively, give
\begin{align*}
  F(z) & = \Phi(z)-n^{-1/2}\dfrac{z^2+2}{6}\phi(z)-n^{-1}\dfrac{2z^5-11z^3+57z}{144}\phi(z)
         + O\big(n^{-3/2}\big)\,, \\
  F^*(z) & = \Phi(z)-n^{-1/2}\dfrac{z^2-1}{6}\phi(z)-n^{-1}\dfrac{2z^5-23z^3+75z}{144}\phi(z)
           + O\big(n^{-3/2}\big)\, ,
\end{align*}
where $\Phi(z)$ and $\phi(z)$ are the distribution and density
functions of the standard normal distribution.  The corresponding
Cornish-Fisher expansions \citep[\S~2.5]{hall92} of the $\alpha$-level
quantiles $q_\alpha$ of $F(z)$ and $q^*_\alpha$ of $F^*(z)$ in terms
of the $\alpha$-level standard normal quantiles $z_\alpha$ are, then,
\begin{align}
  \label{qa}
  q_\alpha & = z_\alpha+ n^{-1/2}\dfrac{z_\alpha^2+2}{6} - n^{-1}\dfrac{11z_\alpha^3-65z_\alpha}{144}+O\big(n^{-3/2}\big)\,, \\
  \label{qsa}
  q^*_\alpha & = z_\alpha + n^{-1/2}\dfrac{z_\alpha^2-1}{6} - n^{-1}\dfrac{11z_\alpha^3-65z_\alpha}{144}+O\big(n^{-3/2}\big)\,,
\end{align}
provided that $\epsilon < \alpha < 1- \epsilon$ for any
$0 < \epsilon < 1/2$. Both $q^*_\alpha$ and $q_\alpha$ have non-zero
$O(n^{-1/2})$ terms, and hence $t^*$ does not deliver an improvement
in the sense of a drop in the asymptotic order of the distributional
approximation. Nevertheless, a careful term-by-term comparison reveals
that
$(q_\alpha - z_\alpha) \simeq (q^*_\alpha - z_\alpha) +
n^{-1/2}/2$. As a result, the quantiles of the distribution of $t^{*}$ are
closer to the standard normal quantiles than those of $t$.

\section{Confidence intervals based on location-adjusted statistics}
\label{sec:confint}
\begin{figure}
  \caption{Computation of the $95\%$ confidence intervals based on
    $t^*$ in Table~\ref{covWald}, using linear interpolation of the
    values of $T(\hat\btheta; \psi) - B(\hat\btheta; \psi)$ on a
    equispaced grid of $20$ values for $\psi$. The solid grey line is
    the linear interpolator, the dotted lines are at $\pm z_{0.975}$,
    and the vertical dashed lines are at the endpoints of the
    intervals.}
  \includegraphics[width = \textwidth]{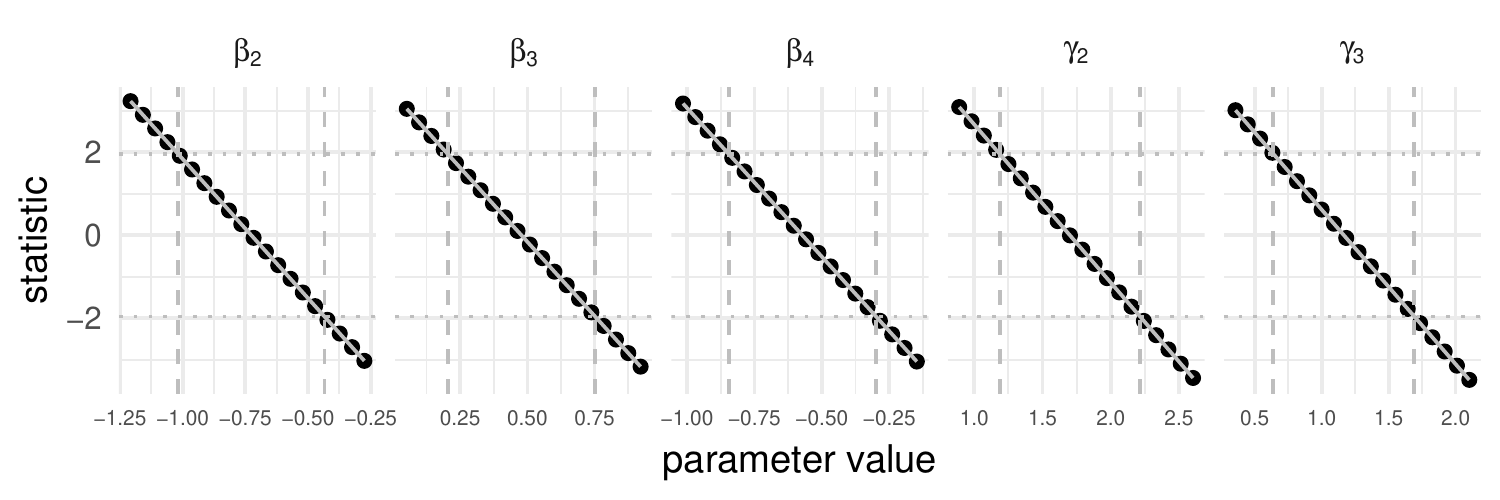}
  \label{interpolation}
\end{figure}

The LA Wald statistics $t^*$ and $\tilde{t}^*$ can be
used to obtain $100(1-\alpha)\%$ CIs by the numerical
inversion of the approximate probability statements for each, that is
by finding all $\psi$ such that
\begin{equation}
  \label{confint}
  z_{\alpha/2} \le  T(\hat\btheta; \psi) - B(\hat\btheta; \psi)  \le z_{1 - \alpha/2}
  \quad \text{and} \quad z_{\alpha/2} \le T(\tilde\btheta; \psi) -
  \tilde{B}(\tilde\btheta; \psi) \le z_{1 - \alpha/2} \, ,
\end{equation}
respectively.
The numerical inversion can be
performed by evaluation of the LA statistics on a grid
of values for $\psi$ and linear
interpolation. Figure~\ref{interpolation} illustrates this process for
the $t^*$ intervals in Table~\ref{covWald}.

\section{Location-adjusted statistics in beta regression}
\label{sec:beta}

\subsection{Implementation}
The LA statistics $\tilde{t}$ and $\tilde{t}^*$ involve
the derivatives of the Wald transform, the expected information matrix
and the first-order bias term of the ML estimator for
beta regression models, whose general expressions, despite being
obtainable, are complicated to either write down or
implement. \citet[\S~2]{grun} provide closed-form expressions for the
latter two, and the \texttt{enrichwith} R package \citep{enrichwith}
can generate the corresponding R functions of the parameters for their
evaluation, which can then be used as input in
Algorithm~\ref{waldi}. In particular, the derivatives of the Wald
transform can be computed using expressions~(\ref{der1:t}) and
(\ref{der2:t}), where $\nabla \kappa(\btheta)$ and
$\nabla\nabla^\top \kappa(\btheta)$ are approximated, to
high-accuracy, for each parameter and at the estimates using
Richardson's extrapolation \citep[as implemented, for example, in the
\texttt{numDeriv} R package;][]{numDeriv}.

\subsection{Reading skills}
For the regression parameters in model~(\ref{eq:m}) of
Example~\ref{reading_skills}, Table~\ref{covWald} reports individual
$95\%$ CIs derived from the inversion of $t^*$ and
$\tilde{t}^*$, and their empirical coverage probabilities. The
empirical coverage probabilities are markedly closer to the nominal
level than those of standard Wald CIs based on the
ML estimates in Table~\ref{tab:fit1}. The intervals
based on $\tilde{t}^*$ are also quite similar to the ones based on
$\tilde{t}$ in Table~\ref{tab:fit1}, with only a slight improvement in
terms of empirical coverage.  This indicates that the extra term
(\ref{biast*br}) in the bias of the Wald statistic based on
RB estimators is of no consequence here; most of the
correction to the location of the Wald statistic is achieved by using
the RB estimator in the place of the ML one.

Table~\ref{covWald} also shows results about the studentized bootstrap
CIs \citep[see,][\S~2.4]{davhin97} for the model
parameters, based on a parametric bootstrap of size $500$ to estimate
the quantiles of the distribution of $t^*$ and $\tilde{t}^*$. The
computation of the studentized bootstrap CIs is done,
again, by numerically inverting the approximate probability statements
for each parameter as in~(\ref{confint}), after replacing the
standard normal quantiles with the corresponding ones estimated via
bootstrap. It should be noted here that in cases like beta regression
an estimator of the cumulative distribution function of the test
statistic being studentized is generally not available in closed
form. As a result, the calculation of studentized bootstrap intervals
is a computationally intensive process. For example, on a MacBook Pro
laptop with 3.5 GHz Intel Core i7 processor and 16 GB of RAM,
computing all CIs based on $\tilde{t}^*$ using
the default arguments in the generic implementation of the
\texttt{waldi\_confint} function in the \texttt{waldi} R package takes
about $1.4$ seconds. On the other hand, the computation of the
corresponding studentized bootstrap intervals takes about $140$
seconds, because of the need to refit the model and calculate the
LA Wald statistics $500$ times.

Using the quantiles of the bootstrap distribution of $t^*$ and
$\tilde{t}^*$ instead of $N(0,1)$ ones, typically brings the empirical
coverage probabilities of intervals based on the LA
Wald statistics even closer to the respective nominal levels, most
notably for the precision effects $\gamma_2$ and $\gamma_3$.

\begin{table}[t]
  \caption{Empirical coverage probabilities of individual confidence
    intervals at nominal levels 90\%, 95\% and 99\% for the regression
    parameters in~(\ref{eq:m}), based on the inversion of the
    location-adjusted Wald statistics $\tilde{t}$ and $\tilde{t}^*$,
    using standard normal quantiles and the quantiles of the bootstrap
    distribution of the statistics from a parametric bootstrap of size
    $500$.  Reported rates are based on $50\, 000$ samples simulated
    under the ML fit in Table~\ref{tab:fit1}.}
  \begin{center}
    \begin{small}
    \begin{tabular}{lrrrrr@{\hskip 2em}cccccc}
      \toprule
& \multicolumn{4}{c}{$95\%$ Confidence intervals} & & \multicolumn{6}{c}{Empirical coverage probability} \\ \midrule
& \multicolumn{2}{c}{$t^*$} & \multicolumn{2}{c}{$\tilde{t}^*$} & & \multicolumn{3}{c}{$t^*$} & \multicolumn{3}{c}{$\tilde{t}^*$} \\ \cmidrule{2-5} \cmidrule{7-12}
& Lower & Upper & Lower & Upper  & & \multicolumn{1}{c}{$90\%$} & \multicolumn{1}{c}{$95\%$} & \multicolumn{1}{c}{$99\%$} & \multicolumn{1}{c}{$90\%$} & \multicolumn{1}{c}{$95\%$} & \multicolumn{1}{c}{$99\%$}\\ \midrule
\multicolumn{12}{c}{Intervals using $N(0, 1)$ quantiles} \\ \midrule
$\beta_2$ & -1.019 & -0.435 & -1.031 & -0.446 & & 88.5 & 93.7 & 98.4 & 88.3 & 93.5 & 98.3  \\
$\beta_3$ &  0.204 & 0.752 & 0.165 & 0.719 & &  87.1 & 92.8 & 98.0 & 87.3 & 93.0 & 98.0 \\
$\beta_4$ & -0.845 & -0.299 & -0.809 & -0.257 & & 87.2 & 92.8 & 98.0 & 87.5 & 93.0 & 98.0 \\
$\gamma_2$ & 1.186 & 2.214 & 1.134 & 2.169 & & 83.5 & 90.0 & 96.6 & 83.9 & 90.3 & 96.8 \\
$\gamma_3$ & 0.639 & 1.691 & 0.513 & 1.574 & & 81.8 & 88.6 & 95.7 & 82.7 & 89.2 & 96.2 \\   \midrule
\multicolumn{12}{c}{Studentized bootstrap intervals} \\ \midrule
$\beta_2$ & -1.059 & -0.442 & -1.091 & -0.440 & & 89.5 & 94.5 & 98.7 & 89.4 & 94.6 & 98.6\\
$\beta_3$ & 0.171 & 0.792 & 0.159 & 0.758 & & 89.2 & 94.3 & 98.5 & 89.4 & 94.5 & 98.5 \\
$\beta_4$ & -0.871 & -0.268 & -0.853 & -0.264 & & 89.3 & 94.3 & 98.5 & 89.5 & 94.4 & 98.6 \\
$\gamma_2$ & 1.112 & 2.303 & 1.040 & 2.241 & & 89.9 & 94.7 & 98.7 & 90.1 & 94.9 & 98.8 \\
$\gamma_3$ & 0.565 & 1.835 & 0.394 & 1.769 & & 90.1 & 94.9 & 98.7 & 90.5 & 95.1 & 98.8 \\ \bottomrule
    \end{tabular}
  \end{small}
\end{center}
  \label{covWald}
\end{table}


\section{Inference about log-odds and binomial proportions}
\label{sec:bernoulli}
Hypothesis tests and confidence intervals about log-odds and binomial
proportions are amongst the most common statistical tasks in applied
data analysis. In this section, we investigate the performance of the
LA Wald statistics in these contexts, and contrast it to
that of classical proposals in the literature.

Suppose that $Y_i$ has a Bernoulli distribution with mean
$\mu_i=e^{\theta}/\big(1+e^\theta\big) \in (0,1)$, $\theta \in
\Re$. The log-likelihood function about $\theta$, the ML estimate, its first-order bias and the expected information are
$l(\theta) = n\bar{y}\theta-n\log\big(1+e^\theta\big)$,
$\hat{\theta}=\log \big\{\bar{y}/(1-\bar{y})\big\}$, $b(\theta)= -\big(1+e^\theta\big)\big(1-e^\theta\big)\big(2ne^\theta\big)^{-1}$ and
$i(\theta)= ne^\theta/\big(1+e^\theta\big)^2$, respectively. The
well-known RB estimator of the log-odds $\theta$ is
$\tilde{\theta} = \log \big\{(\bar{y} + a)/(1-\bar{y} + a)\big\}$,
where $a = n^{-1}/2$ \citep{haldane55, anscombe56}.
The first summand in the right-hand side of (\ref{biastj}) equals
\[ b(\theta)\frac{d T(\theta; \theta_{0})}{d\theta} =  \frac{e^{2\theta}(\theta_0-\theta+2)+2e^\theta(\theta-\theta_0)+\theta_0-\theta-2}{4n^{1/2}e^{\theta/2}(1+e^\theta)}\,,\]
and the second is
\[ \tilde{B}(\theta; \theta_0)= \frac{1}{2i(\theta)}\frac{d^2 T(\theta; \theta_{0})}{d\theta^2}=\frac{e^{2\theta}(\theta-\theta_0-4)+6e^\theta(\theta_0-\theta)+\theta-\theta_0+4}{8n^{1/2}e^{\theta/2}(1+e^\theta)}\,.\]
Hence, we obtain $B(\theta;\theta_0)=(8n^{1/2})^{-1}
(\theta_0-\theta)(e^{-\theta/2}+e^{\theta/2})$. The Wald statistics $t$ and
$\tilde{t}$ and their LA versions are
\begin{align*}
  t  =  &\,\big\{n\big(\bar{y}-\bar{y}^2\big)\big\}^{1/2}
      \bigg(\log\frac{\bar{y}}{1-\bar{y}}-\theta_0 \bigg)\, , \\
  \tilde{t}  =  &\,\frac{\{n(\bar{y}+a)(1-\bar{y}+a)\}^{1/2}}{1+2a}
  \bigg(\log\frac{\bar{y}+a}{1-\bar{y}+a}-\theta_0 \bigg)\, ,\\
    t^{*}  =  &\,\left\{(n\bar{y}-n\bar{y}^2)^{1/2} + (n\bar{y}-n\bar{y}^2)^{-1/2}/8\right\}
          \left( \log \dfrac{\bar{y}}{1-\bar{y}}-\theta_0\right) \, , \\
  \tilde{t}^{*}  =  &\,\,\tilde{t} - \frac{(1-\bar{y}+a)^{3/2}}{8n^{1/2}(\bar{y}+a)^{1/2}(1+2a)} \bigg\{
  \left(\frac{\bar{y}+a}{1-\bar{y}+a}\right)^2 \bigg(\log\frac{\bar{y}+a}{1-\bar{y}+a}-\theta_0 - 4 \bigg) \\
  & \,\,\,\,+ \frac{6(\bar{y}+a)}{1-\bar{y}+a} \bigg(\theta_0 - \log\frac{\bar{y}+a}{1-\bar{y}+a} \bigg)
  + \log\frac{\bar{y}+a}{1-\bar{y}+a}-\theta_0 + 4 \bigg\}\, ,
\end{align*}
respectively. All statistics depend on values of $n\overline{Y}$,
which has a null binomial distribution with index $n$ and probability
$e^\theta/(1 + e^\theta)$. If $\bar{y} = 0$, then
$\hat{\theta}=-\infty$ and if $\bar{y}=1$, then
$\hat{\theta}= +\infty$. In these cases, $t = 0$ and
$t^* = \pm\infty$, respectively. So, regardless of the value
$\theta_0$, when all observations are equal to zero or one the Wald
test always accepts the hypothesis $\theta = \theta_0$, and its
LA version always rejects it. For this reason and to
enable comparison, the convention that $t^* = t = 0$ whenever
$\bar{y}=0$ or $\bar{y}=1$ is adopted for every $\theta_0 \in \Re$. No
such convention is necessary for $\tilde{t}$ and $\tilde{t}^*$ because
the finiteness of $\tilde{\theta}$ guarantees their finiteness.

\begin{figure}[t]
  \caption{The function $\Phi^{-1}(G(z)) - z$ when $G(z)$ is the null
    distribution function for $t$, $t^*$, $\tilde{t}$ and
    $\tilde{t}^*$ under the Bernoulli model of
    Section~\ref{sec:bernoulli}, for $\theta_{0} \in \{-2, -1, 0\}$
    and $n \in \{8, 16, 32 \}$. The reference zero line (grey)
    corresponds to $G(z) = \Phi(z)$.}
  \begin{center}
    \includegraphics[width = 0.95\textwidth]{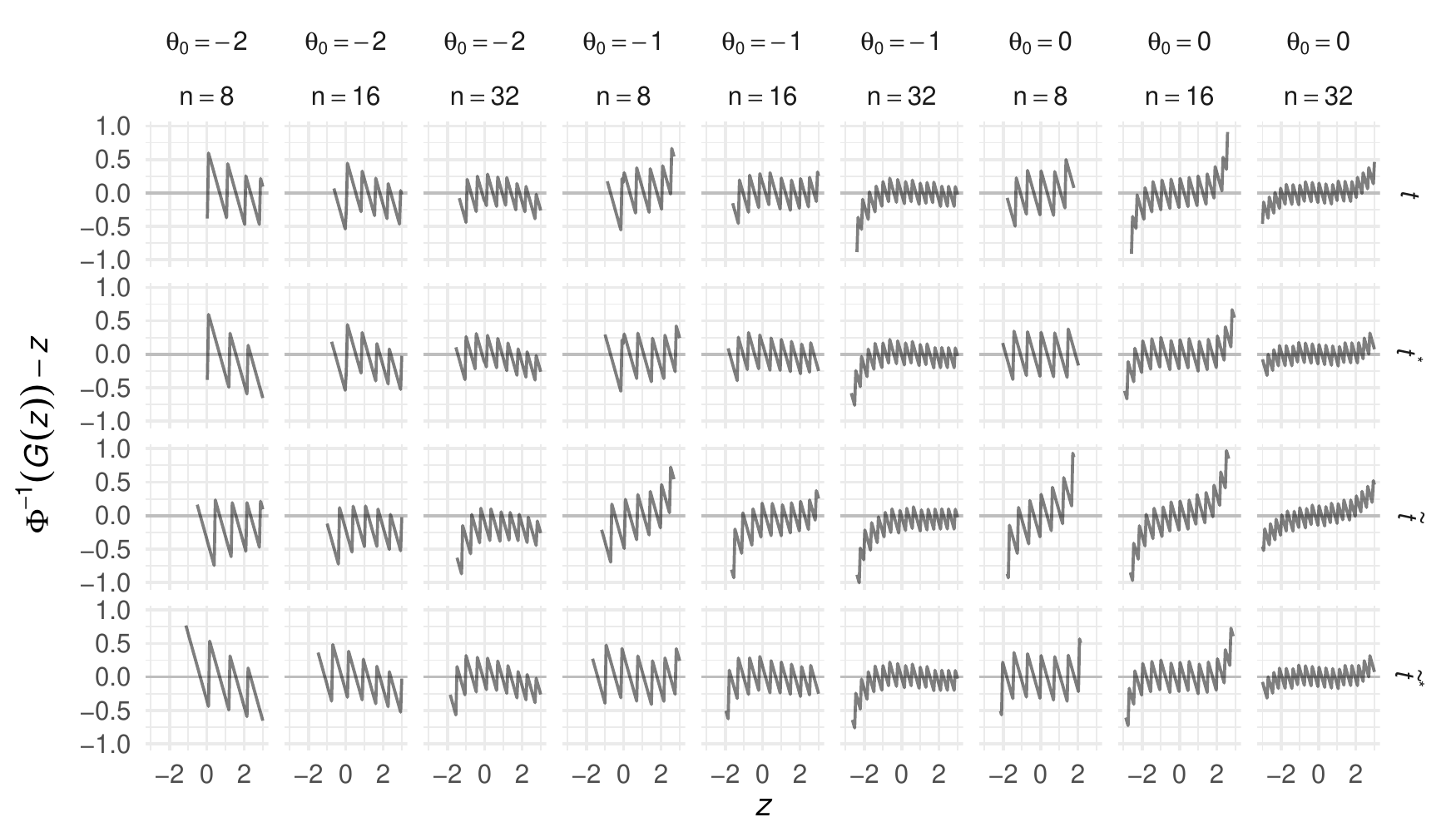}
  \end{center}
	\label{fig:LOGnew2}
\end{figure}

Figure~\ref{fig:LOGnew2} compares $\Phi(z)$ to the null distribution
of $t$ and $t^*$ for $\theta_0 \in \{-2, -1, 0\}$ and
$n\in\{8, 16, 32\}$, plotting only over the range of $z$ where each
distribution takes values in $(0, 1)$. The adjustment of the Wald
statistic appears to be effective in terms of bringing the
corresponding distributions closer to standard normal, especially for
the smaller sample sizes. The statistic $\tilde{t}^*$ performs best
having a distribution that is similar to that of $t^*$, but taking
values in $(0, 1)$ for a wider range of $z$. Note here that, in the
majority of cases, the standard normal approximation to the
distribution of the Wald statistic based on the RB estimator
is worse than that of $t$.

The statistic $\tilde{t}^*$ can be readily inverted numerically as
in~(\ref{confint}) to produce CIs for the log-odds $\theta$ with
excellent coverage properties. Applying the transformation
$e^\theta/(1 + e^\theta)$ to the endpoints of those intervals results
in CIs for the binomial probability. As is demonstrated in the
supplementary material, the latter have competitive coverage
properties and shorter expected lengths, for most values in $(0, 1)$,
than the widely accepted ``add 2 successes and 2 failures'' intervals
proposed in \citet{agresti98} and \citet{agresti00}, with the added
benefit of no overshooting outside $(0, 1)$. Based on these findings,
we propose the transformation of the endpoints of the CIs based on
$\tilde{t}^*$ for routine use when inference about a binomial
proportion or log-odds is of interest.

Finally, the tests conducted via the LA Wald statistic
are found to be more robust to the undesirable behaviour of
Wald tests in binomial settings observed in \citet{haudon}. To
illustrate this, consider $n = 32$ and $\theta_0 = 0$. The values of
the Wald statistic for $\bar{y} = 0.875$, $\bar{y} = 0.906$,
$\bar{y} = 0.938$, $\bar{y} = 0.969$ and $\bar{y} = 1$ are $3.640$,
$3.741$ $3.708$ $3.380$ and $0$ (in 3 significant decimal places),
respectively. As a result, the evidence against $H_0\!: \theta = 0$
from the Wald test decreases when $\bar{y} > 0.906$, despite the fact
that the sample mean is getting further away from the null probability
of $0.5$. In contrast, the corresponding values for the
LA Wald statistic are $3.770$, $3.913$, $3.955$,
$3.816$ and $0$, respectively, thus the evidence against $H_0$
decreases only after $y > 0.938$. The same behaviour is observed for
$\tilde{t}^*$.

\section{Generalized linear models}
\label{sec:glm}

\subsection{Wald statistics}
One of the key summaries in the output of standard statistical
software when fitting generalized linear models are Wald statistics
for the regression coefficients.

In generalized linear models \citep{mcculnel89}, the conditional
distribution of $Y_i$ given $\bx_i$ is assumed to be from the
exponential dispersion family \citep{jorg87}, with density or
probability mass function
\[
  f(y_i | \bx_i; \theta_i) = \exp\left\{\frac{y \theta_i - h(\theta_i) -
      c_1(y)}{\phi/m_i} - \frac{1}{2}a\left(-\frac{m_i}{\phi}\right) +
    c_2(y) \right\} \,.
\]
In the latter expression, $h(\cdot)$, $c_1(\cdot)$, $a(\cdot)$ and $c_2(\cdot)$ are
sufficiently smooth functions, and $m_1, \ldots, m_n$ are known,
non-negative observation weights. Special distributions with density
or probability mass function of the above form are the normal, gamma,
Poisson and binomial.

The conditional expectation $\mu_i = h'(\theta_i)$ of $Y_i$ is linked
to $\bx_i$ as $g(\mu_i) = \eta_i = \bbeta^\top \bx_i$, where $g(\cdot)$ is an
at least three times differentiable link function, and
$h'(u) = d h(u)/du$. The variance of $Y_i$ is $\phi V(\mu_i) /m_i$,
where $V(\mu_i) = h''(\theta_i)$ is the variance function, with
$h''(u) = d^2 h(u)/du^2$, and $\phi$ is a dispersion parameter that
allows shrinking or inflating the contribution of the mean to the
sample variance.

The gradient of the log-likelihood with respect to $\bbeta$ is
inversely proportional to $\phi$ and, hence, the ML
estimate for $\bbeta$ can be obtained without knowing $\phi$, through
iteratively reweighted least squares (IWLS; \citealt{green84}). The expected
information matrix on $\bbeta$ and $\phi$ is
\begin{equation}
  \label{eq:info}
 \bi(\bbeta, \phi)= \left[
    \begin{array}{cc}
      \frac{1}{\phi} \bX^\top \bW(\bbeta) \bX & 0_k \\
      0_k^\top & \frac{1}{2\phi^4}\sum_{i = 1}^n m_i^2 a''(-m_i/\phi)
    \end{array}
  \right]\,,
\end{equation}
where $0_k$ is a $k$-dimensional vector of zeros,
$a''(u) = d^2 a(u)/d u^2$, $\bX$ is the $n \times k$ model matrix with
rows $x_1, \ldots, x_n$ and
$\bW = {\rm diag}\left\{w_1, \ldots, w_n\right\}$ with
$w_i = m_i d_i^2/V(\mu_i)$, $d_i = d\mu_i/d\eta_i$.

The Wald statistic in~(\ref{testf}) that is typically reported for
$H_0:\beta_j= \beta_{j0}$ $(j=1,\dots,k)$ in the output of software
for fitting generalized linear models has the form
\begin{equation}
  \label{tglm}
  t_j = (\hat\beta_j - \beta_{j0})/\kappa_j(\hat\bbeta, \phi^*)
\end{equation}
(see, for example, the \texttt{summary.glm} method in R), and is the
estimate of the Wald transform
$T_j(\bbeta, \phi; \beta_{j0}) = (\beta_j -
\beta_{j0})/\kappa_j(\bbeta, \phi)$, where $\kappa_j(\bbeta, \phi)$
denotes the $(j, j)$th element of the matrix
$[\phi \{\bX^\top \bW(\bbeta) \bX\}^{-1}]^{1/2}$, and $\phi^*$ is an
estimator of $\phi$.

As discussed by \citet{mcculnel89}, the ML estimator
$\hat\phi$ is severely biased and not robust under mis-specification
of the conditional distribution of $Y_i$ given $\bx_i$.  For this
reason, when $\phi$ is unknown, \citet{mcculnel89} recommend to
replace $\phi^*$ in (\ref{tglm}) with the moment estimator
$\tilde\phi = \sum_{i = 1}^n (y_i - \hat\mu_i)^2/\big\{(n -
p)V(\hat\mu_i)\big\}$ that is based on the Pearson residuals.

\subsection{Implementation}
\label{sec:computation_glms}
The bias terms $B_{j}(\bbeta, \phi; \beta_{j0})$ and
$\tilde{B}_{j}(\bbeta, \phi; \beta_{j0})$ can be readily computed for
all generalized linear models using the expression for the first term
in the bias expansion of the ML estimators in
\citet{cordmc91}, the derivatives of $T_{j}(\bbeta, \phi; \beta_{j0})$
in (\ref{der1:t}) and (\ref{der2:t}), the expected information
matrix~(\ref{eq:info}), and its derivatives. The derivatives of
$\bi(\bbeta, \phi)$ with respect to $\bbeta$ can be written in closed form as
\begin{equation}
  \label{eq:der1b}
  \frac{\partial \bi(\bbeta,\phi)}{\partial \beta_u}=\left[
    \begin{array}{cc}
      \frac{1}{\phi} \bX^\top \bW'_u(\bbeta) \bX & 0_k \\
      0_k^\top & 0
    \end{array}
  \right] \quad \text{and} \quad \frac{\partial^2
    \bi(\bbeta,\phi)}{\partial \beta_u\partial \beta_v}=\left[
    \begin{array}{cc}
      \frac{1}{\phi} \bX^\top \bW''_{uv}(\bbeta) \bX & 0_k \\
      0_k^\top & 0
    \end{array}
  \right]\,,
\end{equation}
where $\bW'_u= \bW(2\bR-\bL)T_u$, with $\bR=\mathrm{diag}\{r_{1},\dots,r_{n}\}$,
$r_i=d \log d_i/d\eta_i$, $\bL=\mathrm{diag}\{l_{1},\dots,l_{n}\}$,
$l_i= d \log V(\mu_i)/d\eta_i$, and
$\bT_u=\mathrm{diag}\{x_{1u},\dots,x_{nu}\}$. Furthermore,
$\bW''_{uv}=\bW(2\bR-\bL)^2\bT_u\bT_v + \bW(2\bR'-\bL')\bT_u\bT_v$, with
$\bR'=\mathrm{diag}\{r'_{1},\dots,r'_{n}\}$,
$r'_i = d^2 \log d_i /d\eta^2_i$ and
$\bL'=\mathrm{diag}\{l'_{1},\dots,l'_{n}\}$,
$l'_i= d^2 \log V(\mu_i)/d\eta^2_i$ $(u,v = 1, \ldots, k)$. For
generalized linear models with unknown dispersion parameter, the
derivatives of $\bi(\bbeta, \phi)$ with respect to $\phi$ are
\begin{equation}
  \label{eq:der1p}
  \frac{\partial \bi(\bbeta,\phi)}{\partial \phi}=\left[
    \begin{array}{cc}
      -\frac{1}{\phi^2} \bX^\top \bW(\bbeta) \bX & 0_k \\
      0_k^\top & \frac{1}{2\phi^6}\sum_{i = 1}^n m_i^2 a'''(-m_i/\phi)-\frac{2}{\phi^5}\sum_{i = 1}^n m_i^2 a''(-m_i/\phi)
    \end{array}
  \right]
\end{equation}
and
\begin{equation}
  \label{eq:der2p}
  \frac{\partial^2 \bi(\bbeta,\phi)}{\partial \phi^2}=\left[
    \begin{array}{cc}
      \frac{2}{\phi^3} \bX^\top \bW(\bbeta) \bX & 0_k \\
      0_k^\top & \frac{10}{\phi^6}\sum_{i = 1}^n m_i^2 a''(-m_i/\phi)-\frac{5}{\phi^7}\sum_{i = 1}^n m_i^2 a'''(-m_i/\phi) \\ & +\frac{1}{2\phi^8}\sum_{i = 1}^n m_i^2 a^{iv}_i(-m_i/\phi)
    \end{array}
  \right]\,,
\end{equation}
and the mixed second derivative is
\begin{equation}
  \label{eq:derbp}
  \frac{\partial^2 \bi(\bbeta,\phi)}{\partial \beta_u\partial \phi}=\left[
    \begin{array}{cc}
      -\frac{1}{\phi^2} \bX^\top \bW'_{u}(\bbeta) \bX & 0_k \\
      0_k^\top & 0
    \end{array}
  \right] \, .
\end{equation}

Algorithm~\ref{waldi} for computing the LA statistics can also be
implemented by replacing steps~15 and 16 with the evaluation of
derivatives~(\ref{eq:der1b}), (\ref{eq:der1p}), (\ref{eq:der2p}) and
(\ref{eq:derbp}). The \texttt{waldi} R package can compute the LA
statistics using either the analytical derivatives or numerical
differentiation of $\kappa(\bbeta, \phi)$, as in
Section~\ref{sec:beta}.

\subsection{Gamma regression with unknown dispersion}
\label{sec:clot}

\begin{table}[t]
  \caption{The ML estimates of the regression
    parameters $\beta_1$, $\beta_2$, $\beta_3$ and $\beta_4$ of the
    gamma regression model for the blood clotting dataset, and the
    ML ($\star$) and moment-based ($\dagger$) estimate of
    $\phi$. The various versions of the Wald statistic are for the
    individual hypotheses $H_0: \beta_{j} = 0$ $(j=1,\dots,4)$.}
  \begin{center}
    \begin{small}
    \begin{tabular}{crrrr}
      \toprule
      & & \multicolumn{2}{c}{Wald} & \\ \cmidrule{3-4}
      & \multicolumn{1}{c}{Estimate} & \multicolumn{1}{c}{$\hat{\phi}$} & \multicolumn{1}{c}{$\tilde{\phi}$} & \multicolumn{1}{c}{$t^*$} \\
      \midrule
      $\beta_1$ & 5.503 & 34.126 & 29.282 & 28.953 \\
      $\beta_2$ & -0.602 & -12.842 & -11.020 & -10.896\\
      $\beta_3$ & -0.584 & -2.563 & -2.199 & -2.173\\
      $\beta_4$ & 0.034  & 0.520 & 0.446 & 0.441\\
      $\phi$ & $^\star$0.017 \\
      & $^\dagger$0.024  \\
      \bottomrule
    \end{tabular}
  \end{small}
\end{center}
  \label{tab:clotFIT}
\end{table}

The data in \citet[\S~8.4.2]{mcculnel89} consist of 18 observations of
mean blood clotting times in seconds for nine percentage
concentrations of normal plasma and two lots of clotting agent. The
clotting times are assumed here to be realizations of independent
random variables $Y_1, \ldots, Y_{18}$, each $Y_i$ having a
Gamma$(\phi^{-1}, (\phi\mu_i)^{-1})$ distribution with
\[
  \log\mu_i=\beta_1+\sum_{j=2}^4\beta_jx_{ij}\quad (i = 1, \dots, 18)\,,
\]
where $x_{i2}$ denotes the normal plasma concentration, $x_{i3}$ is
a dummy variable encoding which of the two lots of clotting agent was
employed for the $i$th observation, and $x_{i4} = x_{i2}x_{i3}$ is the
interaction between the plasma concentration and the clotting agent.

\begin{figure}[t]
  \caption{Empirical null rejection probabilities when testing
    $H_0:\beta_j = \beta_{j0}$ against $H_1:\beta_j \ne \beta_{j0}$
    (top row) and $H_1\!: \beta_j < \beta_{j0}$ (bottom row)
    $(j = 2, 3, 4)$ based on the normal approximation to the
    distribution of $t^*$ (triangles) and the Wald statistic using
    $\tilde\phi$ (squares) and $\hat\phi$ (circles). The null value
    $\beta_{j0}$ is set at the estimate of $\beta_j$ in
    Table~\ref{tab:clotFIT}. Reported rates obtained using $50\, 000$
    simulated samples from the ML fit shown in
    Table~\ref{tab:clotFIT}, and for nominal levels (dashed grey line)
    $0.1\%$ (left-most column), $1\%$ (second column from left),
    $2.5\%$ (third column from left), and $5\%$ (right column).}
  \begin{center}
    \includegraphics{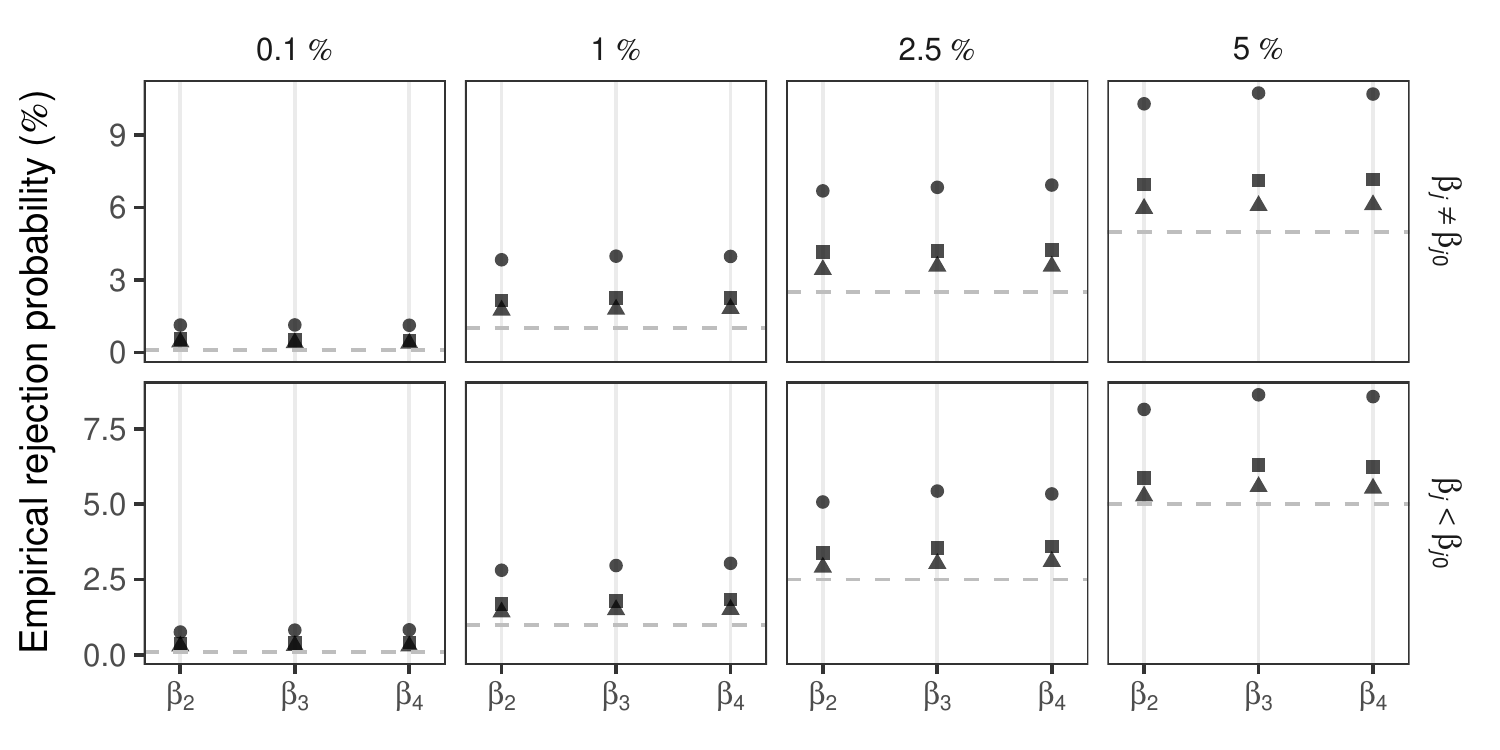}
  \end{center}
  \label{fig:clottingsimu}
\end{figure}

Table~\ref{tab:clotFIT} reports the ML estimates of
the regression parameters and the Wald statistics for the individual
hypotheses $H_0: \beta_{j}=0$ $(j=1,\dots,4)$ using either $\hat \phi$
or $\tilde \phi$ in~(\ref{tglm}), along with the corresponding
LA Wald statistics. The values of the Wald statistic
with $\tilde\phi$ are closer to $t_j^*$ than those with $\hat\phi$,
confirming that the recommendation of \citet{mcculnel89} to use
$\tilde\phi$ in (\ref{tglm}) delivers some location
correction. Figure~\ref{fig:clottingsimu} shows empirical null
rejection probabilities at several nominal levels when testing
$H_0:\beta_j = \beta_{j0}$ against $H_1:\beta_j \ne \beta_{j0}$ and
$H_1:\beta_j < \beta_{j0}$ $(j = 2, 3, 4)$, where $\beta_{j0}$ is set
at the estimate of $\beta_j$ in Table~\ref{tab:clotFIT}. The test
based on the LA Wald statistic performs better than
that based on the Wald statistic with $\hat\phi$ and $\tilde\phi$. As
expected, the latter is the least reliable. The statistics
$\tilde{t}_j$ and $\tilde{t}^*_j$, based on RB estimators of
$\bbeta$ and $\phi$, result in the same, to plotting accuracy,
rejection probabilities as the Wald statistic using $\tilde\phi$ and
$t^*_j$, respectively (see supplementary material). For this reason,
we have not reported their values in Table~\ref{tab:clotFIT} and
empirical rejection probabilities in Figure~\ref{fig:clottingsimu}.

\subsection{Logistic regression with many nuisance parameters}
\label{sec:babies}

The dataset in \citet[Table~1]{dav88} records 18 matched pairs of
binomial observations from a study that investigates the effect of lulling on the
crying of babies. Matching is per day and each day pair consists of
the number of babies not crying out of a fixed number of control
babies, and the outcome of ``lulling'' on a single child. A total of
$143$ babies are involved in the experiment. Interest is in testing
the effect of lulling on the crying of children. Suppose that the
crying status of baby $j$ in day $i$ is a Bernoulli random variable
with probability $\mu_{ij}$ of not crying such that
\begin{equation}
  \label{crying_babies}
  \log\frac{\mu_{ij}}{1 - \mu_{ij}} = \beta_{i} + \gamma z_{ij} \quad
  (i = 1, \ldots, 18;\,j = 1, \ldots, n_i)\,,
\end{equation}
where $z_{ij}$ takes value one if the $j$th baby out of the $n_i$
children observed on day $i$ was lulled, and zero otherwise. We assume
independence between babies and across days. This generalized linear
model has $\phi = 1$, and $18$ nuisance parameters to allow for
different probabilities of not crying across days and account for
between-day variations in the experimental design. Estimation and
inference about $\gamma$ can be performed via either the likelihood
or the conditional likelihood after eliminating
$\beta_1, \ldots, \beta_{18}$ by conditioning on their sufficient
statistics (see, for example, \citeauthor{agr02},
\citeyear[\S~6.7.1]{agr02}).

\begin{table}[t]
  \caption{Maximum likelihood ($\hat\gamma$), maximum conditional
    likelihood ($\hat\gamma_c$) and reduced-bias $\tilde\gamma$
    estimates for $\gamma$ in~(\ref{crying_babies}), with
    corresponding estimated standard errors (in parenthesis). The
    statistics are for $H_0\!: \gamma = 0$ and involve the Wald
    statistic using the ML, maximum conditional
    likelihood and reduced-bias estimates ($t$, $t_c$ and $\tilde{t}$,
    respectively), the signed roots of the logarithms of the
    likelihood and conditional likelihood ratio statistics ($r$ and
    $r_c$, respectively), and the location-adjusted Wald statistics
    based on the ML and reduced-bias estimates ($t^*$
    and $\tilde{t}^*$, respectively). Approximate $p$-values based on
    the normal distribution are given in square brackets.}
  \begin{center}
    \begin{small}
    \begin{tabular}{cccccccccc}
      \toprule
      \multicolumn{1}{c}{$\hat\gamma$} &
                                         \multicolumn{1}{c}{$\hat\gamma_{c}$}
      & \multicolumn{1}{c}{$\tilde\gamma$}& \multicolumn{1}{c}{$t$} &
                                                                      \multicolumn{1}{c}{$t_{c}$} & \multicolumn{1}{c}{$\tilde{t}$} & \multicolumn{1}{c}{$r$} & \multicolumn{1}{c}{$r_c$} & \multicolumn{1}{c}{$t^*$} & \multicolumn{1}{c}{$\tilde{t}^*$} \\
      \midrule
      1.4324 & 1.2561 & 1.1562 & 1.9511 & 1.8307 & 1.7362 & 2.1596 &
                                                                     2.0214
                                                                                                                                                                                          & 1.9257 & 1.9064 \\
      (0.7341) & (0.6861) & (0.6659) & [0.0510] & [0.0671] & [0.0825]
                                                                                                                                    &
                                                                                                                                      [0.0308] & [0.0432] & [0.0541] & [0.0566] \\
      \bottomrule
    \end{tabular}
  \end{small}
\end{center}
  \label{tab:babies}
\end{table}

Table~\ref{tab:babies} reports the ML and maximum conditional
likelihood (MCL) estimates of $\gamma$, the corresponding standard
errors, and the values of typical statistics for testing $\gamma =
0$. The statistics $\tilde{t}$ and $\tilde{t}^*$ use the bias-reducing
adjusted score estimators in \citet{firth93} as implemented in the
\texttt{brglm2} R package \citep{brglm2}. A conventional parametric
bootstrap based on $1000$ samples under the ML fit gives a $p$-value
of $0.0230$, which is similar to what a LR test ($r$ in
Table~\ref{tab:babies}) gives. As is apparent, the values of $t^*$ and
$\tilde{t}^*$ are closer to that of the Wald statistic based on the
MCL estimator $\hat\gamma_c$.

\begin{figure}[t]
  \caption{Empirical null $p$-value distributions when testing
    $H_0\!: \gamma = 0$ against $H_1:\gamma \ne 0$ (top row) and
    $H_1:\gamma < 0$ (bottom row) using the normal approximation to
    the distribution of $t$, $\tilde{t}$, $t_c$, $r$, $r_c$, $t^*$,
    $\tilde{t}^*$ and parametric bootstrap based on 1000
    samples under the ML fit ($boot$). The solid
    horizontal line at one is the uniform density. Results obtained
    from a simulation study with $50\,000$ replications.}
  \begin{center}
    \includegraphics{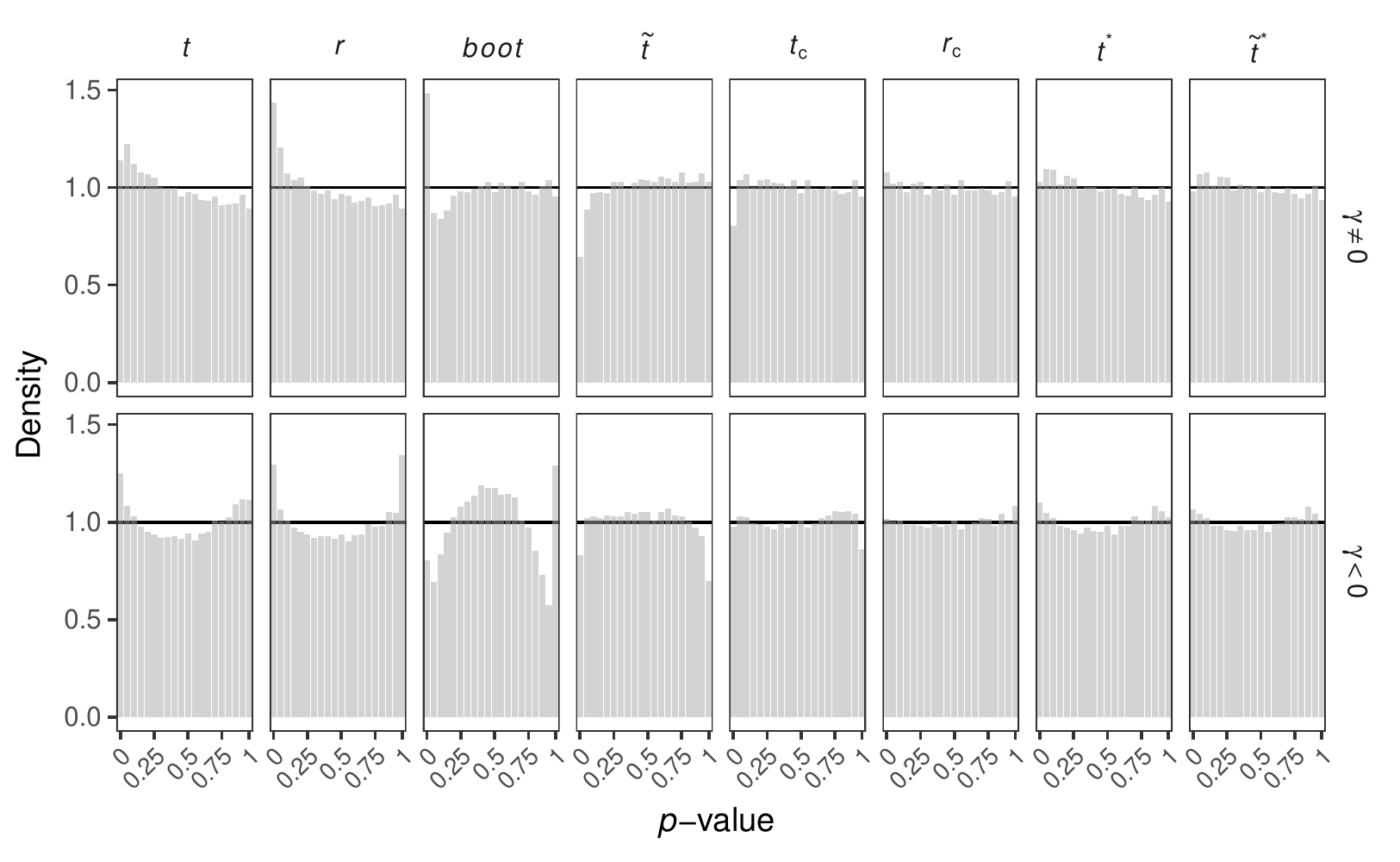}
  \end{center}
  \label{babies_histogram}
\end{figure}

Figure~\ref{babies_histogram} shows the empirical null $p$-value
distribution for the various statistics in Table~\ref{tab:babies} when
testing $H_0:\gamma = 0$ against $H_1:\gamma \ne 0$ and
$H_1:\gamma < 0$. These distributions are computed using $50\, 000$
samples from model~(\ref{crying_babies}) with
$\beta_1, \ldots, \beta_{18}$ set to their ML estimates from the
observed data and $\gamma = 0$. There were $13$ samples where
$\hat\gamma$ and $\hat\gamma_c$ had infinite value and the LA Wald
statistic for $\gamma$ was given value zero, in line with the
discussion in Section~\ref{sec:bernoulli}. No such conventions are
necessary for the computation of $\tilde{t}$ and $\tilde{t}^*$,
because the RB estimator is always finite \citep[see,][for
proof]{kos2018}. The detection of infinite estimates was done
prior to fitting the models using the linear programming algorithms in
the unpublished PhD thesis of \citet{konis07}, as implemented in the
\texttt{detect\_separation} method of the \texttt{brglm2} R package.

The empirical null $p$-value distributions for $t$ and for the signed
root of the logarithm of the LR statistic $r$ are far
from uniform. This is a well-studied issue when testing in the
presence of nuisance parameters \citep{dav88}, which can be remedied
by their elimination via conditioning on their sufficient statistics
and carrying out inference based on the conditional likelihood;
Figure~\ref{babies_histogram} illustrates that the distribution of
$p$-values based on $r_c$ is much closer to uniform than that of $t$,
$r$ and parametric bootstrap based on 1000 samples under the ML
fit. Despite of its simplicity and of being based on a single fit of
the model that involves all nuisance parameters, the LA statistic
$t^*$ and, particularly, $\tilde{t}^*$ deliver a dramatic improvement
over all $t$, $r$, and bootstrap, having $p$-value distributions that
are close to uniform and having significantly lower computational cost
than bootstrap, $r$, and $r_c$.

\section{Significance maps from brain lesion data}
\label{sec:MRI}
\citet{ge2014} propose a spatial probit model and develop the
associated Bayesian machinery for the analysis of multiple sclerosis
lesion maps, accounting for spatial dependence between lesion location
and subject-specific covariates. After warping the brain data for each
individual in the sample to a common atlas, the resulting lesion maps
comprise one binary observation in each voxel, indicating the presence
or absence of a lesion. The alternative approach, which is also
explored in \citet[\S~4.1]{ge2014}, is mass univariate
modelling of lesion occurrence, where a binary-response generalized
linear model
is fitted independently on each voxel. A key inferential output from
either approach are significance maps, which highlight voxels
according to the evidence against the null hypothesis of no covariate
effect.

Spatial probit regression properly accounts for spatial dependence
using a probit model with multivariate conditional auto-regressive
priors for the regression parameters. However, as the resolution of
the brain scans and the number of individuals increases, posterior
sampling becomes a computationally tedious task due to the large
dimension of the parameter space. For example, the moderately sized
application in \citet{ge2014} has 250 individuals with 274\,596 voxels
each, and requires sampling about 2\,750\,000 parameters from the
posterior. \citet{ge2014} achieved this by developing a
partly-parallelizable Gibbs sampling procedure and distributing the
computation using GPUs. On the other hand, mass univariate regression
does not formally account for spatial variation, but is, nevertheless,
computationally attractive. It can be carried out on a regular laptop
in a matter of minutes, because of the concavity of the log-likelihood
and the ability to parallelize computations over the
voxels. Significance maps can then be produced by relying on the voxel
value of standard statistics, like the Wald and the signed root of the
logarithm of the LR statistic. \citet{ge2014} used Wald
statistics from voxel-wise logistic regressions based on the
RB estimator in \citet{firth93}.

\begin{figure}[t]
  \caption{A single sagittal slice of the average white matter map
    overlayed with non-zero counts of lesions per voxel amongst the 50
    individuals (top left), significance maps for disease duration based on
    $\tilde{t}$ and $\tilde{t}^*$ (bottom left and bottom right,
    respectively), and $\tilde{t}$ versus $\tilde{t}^*$ (top
    right). Voxels with statistic values between $-1$ and $1$ are not
    shown.}
  \begin{center}
    \includegraphics[width = 0.4\textwidth]{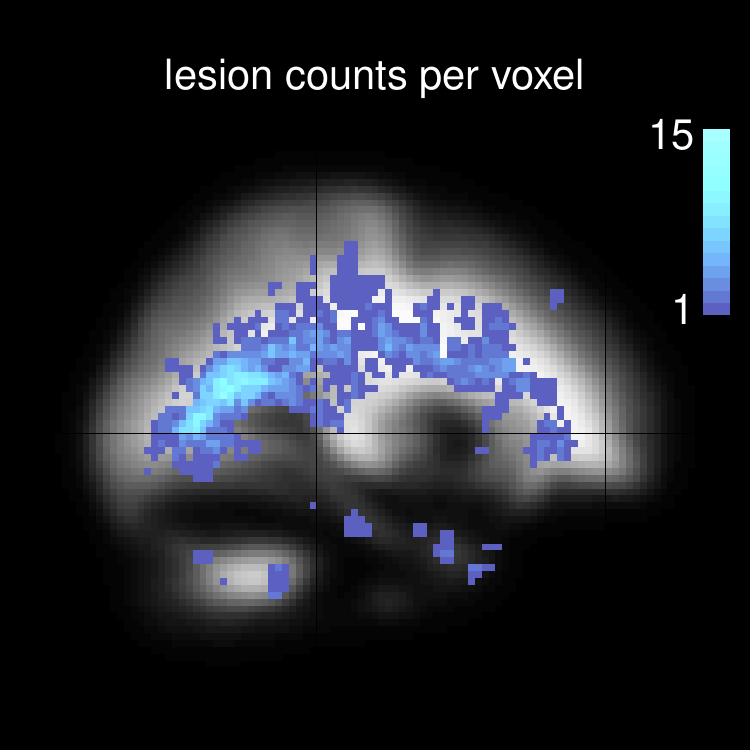}
    \includegraphics[width = 0.4\textwidth]{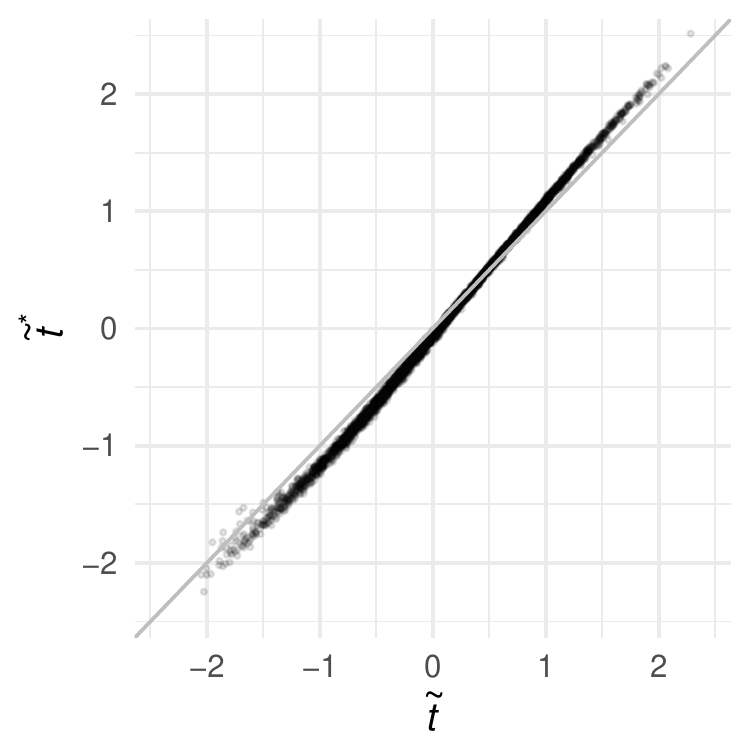}
    \includegraphics[width = 0.4\textwidth]{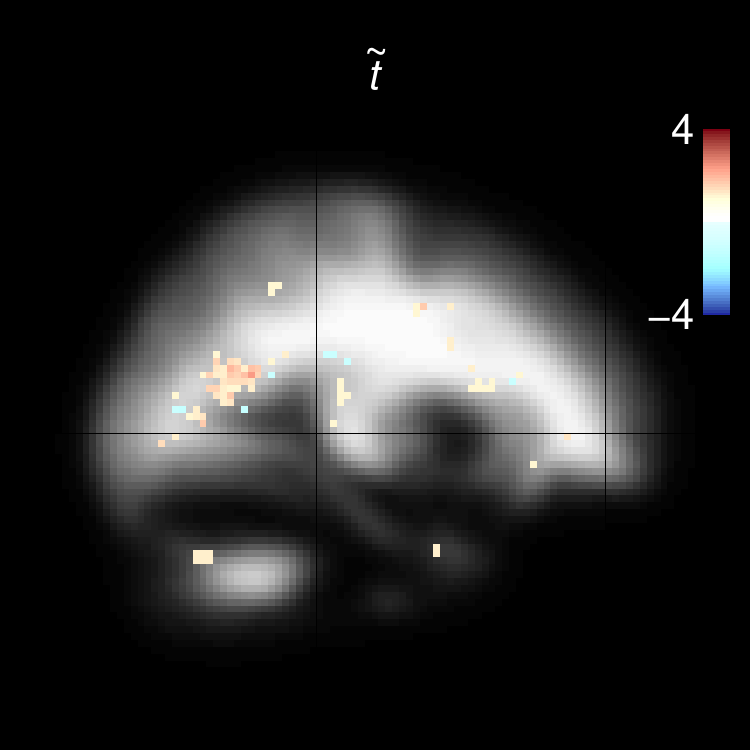}
    \includegraphics[width = 0.4\textwidth]{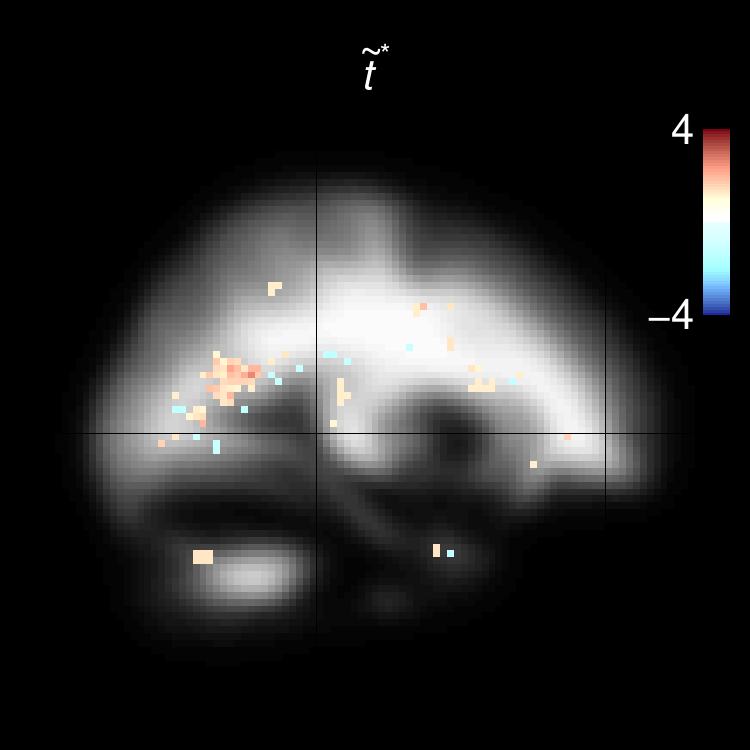}
  \end{center}
\label{fig:brains}
\end{figure}

Here, we use the sample of 50 patients from the supplementary material
of \citet{ge2014} to construct significance maps using both
$\tilde{t}$ and its LA version $\tilde{t}^*$ for the
parameters of a binary-response generalized linear model with probit
link. The available covariates are multiple sclerosis type (0 if
relapsing-remitting multiple sclerosis, 1 if secondary
progressive multiple sclerosis), and standardized versions of age,
sex, disease duration and two disease severity measures
\citep[PASAT and EDSS; see][for details]{ge2014} for each
individual. The lesion maps for the 50 individuals have 902\,629
voxels each (resolution $91 \times 109 \times 91$). Even with such a
small sample of patients, there are $4304$ unique configurations of
lesion occurrence across the voxels in the sample. One of those is the
trivial configuration where there is no lesion occurrence for all
patients, and which happens for 879\,237 voxels. So, for mass
univariate regression we need to fit $4304$ probit regression models,
each with an intercept and 6 parameters, one for each covariate.

The ML estimates for the parameters of age, disease duration, EDSS,
PASAT, sex and type, were infinite for $63.7\%$, $63.7\%$, $63.2\%$,
$63.6\%$, $78.3\%$ and $75.5\%$ of the non-trivial voxels,
respectively, making $t$ and $t^*$ not very useful to produce
significance maps. On the other hand, the computation of $r$ (the
signed root of the logarithm of the LR statistic)
requires $6$ times more fits --- one extra fit of the nested model
which results by omitting each covariate --- than for Wald
statistics. In addition, that computation fails in a considerable
amount of cases due to separation either in the full or the nested
models. This is because standard numerical algorithms (like the IWLS
implementation in the \texttt{glm} R function) reach the maximum
number of iterations without achieving the maximum of the
log-likelihood, resulting in a negative value of the logarithm of the
LR statistic. For example, the calculation of $r$ failed
in $18.1\%$ of non-trivial voxels, while testing for disease duration.

Figure~\ref{fig:brains} shows the results for disease duration
accounting for all other covariates, using a single sagittal slice of
the average white matter map supplied in the supplementary material of
\citet{ge2014}. As expected, larger disease duration corresponds to
more damage from multiple sclerosis, which is apparent by the positive
association between disease duration scores and lesion occurrence,
especially along the minor forceps. Use of $\tilde{t}$ results in
$18.9 \%$ of the voxels being greater than 1 in absolute
value. The corresponding percentage for the LA
$\tilde{t}^*$ is $24.8\%$. As can be seen in the top right plot of
Figure~\ref{fig:brains}, the location adjustment typically results in
an inflation of the value of the Wald statistic, leading, in turn, to
stronger signal detection.

\section{Bootstrap for location- and scale-adjusted statistics}
\label{sec:scale}

A further enhancement to inferential procedures using the Wald
statistic can be achieved by also correcting for its variance. As
in~(\ref{reg:t*}), we can define a location- and scale-adjusted
statistic $t^{**} = (t - \hat{B}) \hat{V}^{-1/2}$, where $\hat{V}$ is
an estimator of the variance $V(\theta, \psi_0)$ of $t$
in~(\ref{reg:t}) or $t^{*}$ in (\ref{reg:t*}). One way to obtain an
estimator for $V(\theta, \psi_0)$ is to derive the series expansion
\[
  V(\theta, \psi_0) = 1 + V_2(\theta, \psi_0) + O(n^{-2}) \, ,
\]
where $V_2(\theta, \psi_0) = O(n^{-1})$, and use
$\hat{V} = 1 + V_2(\hat\theta, \psi_0)$. Section~2.2 of the
unpublished PhD thesis by Claudia Di Caterina \citep{dicaterina17}
evaluates the performance of such a scale adjustment in one-parameter
problems. It is therein found that, while the location- and scale-adjusted Wald
statistic has, in theory, a distribution that is asymptotically
closer to $N(0, 1)$, $\hat{V}$ can take extreme, or even negative,
values in prominent modelling settings. This renders the evaluation
of $t^{**}$ unstable or impossible.

A stable scale adjustment results by estimating $V(\theta, \psi_0)$
from the bootstrap distribution of $t^*$. The computation required is,
practically, the same as the one used for getting studentized
CIs in Section~\ref{sec:beta}, with the difference
that the bootstrap samples are used to estimate a variance instead of
quantiles at tail probabilities. The bootstrap scale adjustment
applies directly to the LA Wald statistic $\tilde{t}^*$
based on RB estimators in Section~\ref{sec:RBest}, to give
the location- and scale-adjusted statistic $\tilde{t}^{**}$.

As an illustration, for the beta regression model of Example
\ref{reading_skills}, the empirical coverage probabilities of $90\%$,
$95\%$ and $99\%$ CIs for the interaction parameter
$\beta_4$ derived by the inversion of $t^{**}$ and $\tilde{t}^{**}$
using $500$ bootstrap samples are $89.8\%$, $94.6\%$, $98.7\%$, and
$90.2\%$, $94.8\%$, $98.8\%$, respectively. For the gamma regression
model in Section~\ref{sec:clot}, the empirical null rejection
probabilities when testing $H_0:\beta_4 = \beta_{40}$ at levels 0.1\%,
1\%, 2.5\%, 5\% via $t^{**}$ based on 500 bootstrap replicates are,
respectively, 0.32\%, 1.50\%, 3.02\%, 5.33\% for
$H_1:\beta_4 \ne \beta_{40}$, and 0.24\%, 1.25\%, 2.69\%, 4.95\% for
$H_1\!: \beta_3 < \beta_{30}$ (see supplementary material for more
extensive results).

While the inversion of $t^{**}$ or $\tilde{t}^{**}$ to get CIs is
possible, it is clearly a laborious process because a separate lot of
bootstrap samples is necessary at each point on the grid of values
considered for $\psi$ (see Section~\ref{confint}). Nevertheless, the
computation of $p$-values based on $t^{**}$ or $\tilde{t}^{**}$
requires only the generation of a single lot of bootstrap samples.

\section{Concluding remarks}
\label{sec:conc}
Correcting the first moment of the distribution of the Wald statistic
has been found to be particularly effective in scenarios where the
bias of the ML estimator impacts inferential
conclusions on the parameter of interest. In contrast to its main
competitors, such as the bootstrap and the LR and score
statistics, the LA Wald statistic has computational
complexity similar to that of the standard Wald one and does not need
any extra model fits. As is done in the algorithm of
Section~\ref{sec:comp}, numerical differentiation can be used to
approximate the gradient and hessian of $\kappa(\btheta)$, in order to
deliver general, modular implementations of the location adjustment
that depend only on a ready implementation of the expected information
matrix and the bias function. The \texttt{waldi} R package in the
supplementary material provides such implementations. This approach,
though general, does not scale as well as analytical derivatives do
with $p$, due to the matrix inversions required during numerical
differentiation. For instance, using numerical derivatives, the
calculation of the LA statistic $t^*$ in
Table~\ref{tab:babies} takes approximately $6$ times the computing
time the analytical calculation does (using the \texttt{waldi} R
package on a MacBook Pro laptop with 3.5 GHz Intel Core i7 processor
and 16 GB of RAM; see supplementary material).

Furthermore, the theoretical framework for LA Wald
statistics is modular, allowing to utilise alternative
variance-covariance matrices or bias estimates. The derivation of the
bias of the Wald statistic in~(\ref{reg:t}) depends only on an
explicit expression for the variance-covariance matrix of the
estimator. Hence, the location adjustment can be performed with more
general forms for the variance-covariance matrix of the asymptotic
distribution of the estimator, including robust versions of it, like
the ones of \citet{mackin85}. Moreover, for $\bbb(\btheta)$
in~(\ref{biastj}), we used the first term in the expansion of the bias
of the ML estimator. In fact, the theory in
Section~\ref{sec:regset} applies unaltered to any approximation of
$\bbb(\btheta)$ that satisfies
$E_{\btheta}(\hat{\btheta} - \btheta) = \bbb(\btheta) +
o\big(n^{-1}\big)$. This included the bias itself, if that is
available, and simulation-based estimators of it, obtained by
jackknifing or parametric and non-parametric bootstrap, as described
in \citet[Chapter 10]{efrtib93} and \citet[Chapter 2]{davhin97}. The
expression of the bias of the Wald statistic in~(\ref{biastj}) is also
the same for Wald statistics based on any estimator that is
$\sqrt{n}$-consistent, asymptotically normal and has a known
variance-covariance matrix.

A bootstrap procedure that exploits the computational simplicity of
LA Wald statistics has also been introduced to deliver
location- and scale-adjusted Wald statistics. Note here that the
variance of $t^*$ can be reliably estimated with less bootstrap
samples than those needed for accurate estimation of quantiles or tail
probabilities of its bootstrap distribution. The latter are what is
required, for instance, to construct studentized CIs
and carry out the bootstrap hypothesis tests.


Further research can surely deal with the theoretical evaluation of the
LA Wald statistic in the stratified settings considered by \cite{sartori03}. In addition,
the quality of the normal approximation to the distribution of the 
location- and scale-adjusted statistics, which were here only briefly outlined,
deserves more detailed investigations \citep[see, e.g.,][for relevant discussions
in the context of bootstrap prepivoting]{lee05}. 
Another direction for future work involves the more systematic
comparison of the various statistics for the construction of
significance maps, including permutation-based approaches \citep[see,
among others,][]{winkler14}.

In closing, the main criticism about Wald procedures is their
lack of invariance under non-linear transformations of the
parameter. Specifically, the conclusions from Wald inferences depend
on the parameterization of the model \citep[see, for instance,][who
tackle this issue by introducing a geometric invariant Wald
statistic]{larjup03}. The LA Wald statistics
improve significantly over standard ones, but their direct
dependence on the Wald statistic and its bias renders them also not
invariant. If parameterization invariance is critical to the inferential problem at
hand, then both Wald statistics and their LA versions should
be used with care.

\section*{Supplementary material}
The supplementary material (also available at
\url{https://github.com/ikosmidis/waldi/tree/master/inst/supplementary_1710-11217})
provides code to fully reproduce all the numerical results and outputs
in the paper, including additional and enriched outputs from the
numerical computations and simulation experiments.

\section*{Acknowledgments}
Ioannis Kosmidis was supported by The Alan Turing Institute under the
EPSRC grant EP/N510129/1 (Turing award number TU/B/000082). Part of
this work was completed when Ioannis Kosmidis was a Senior Lecturer at
University College London, where Claudia Di Caterina spent a year as a
visiting PhD student. The authors are grateful to Thomas Nichols for
helpful discussions and pointers on mass univariate regression for the
occurrence of brain lesions. The authors also thank Nicola Sartori and
Alessandra Salvan for useful talks about the methodology.

\bibliographystyle{chicago}
\bibliography{corzed_2}


\end{document}